\documentclass[sigconf]{acmart}

\AtBeginDocument{%
  \providecommand\BibTeX{{%
    \normalfont B\kern-0.5em{\scshape i\kern-0.25em b}\kern-0.8em\TeX}}}

\copyrightyear{2021}
\acmYear{2021}
\setcopyright{acmlicensed}\acmConference[AIES '21]{Proceedings of the 2021 AAAI/ACM Conference on AI, Ethics, and Society}{May 19--21, 2021}{Virtual Event, USA}
\acmBooktitle{Proceedings of the 2021 AAAI/ACM Conference on AI, Ethics, and Society (AIES '21), May 19--21, 2021, Virtual Event, USA}
\acmPrice{15.00}
\acmDOI{10.1145/3461702.3462617}
\acmISBN{978-1-4503-8473-5/21/05}

\begin{document}
\fancyhead{}

\title{Skilled and Mobile: Survey Evidence of AI Researchers' Immigration Preferences}





\author{Remco Zwetsloot}
\email{Remco.Zwetsloot@georgetown.edu}
\affiliation{%
  \institution{Center for Security and Emerging Technology, Georgetown University}
  \streetaddress{3700 O St NW}
  \city{Washington}
  \state{DC}
  \country{USA}
  \postcode{20057}
}

\author{Baobao Zhang}
\affiliation{%
  \institution{Department of Government, Cornell University}
  \city{Ithaca}
  \state{NY}
  \country{USA}
  \postcode{14853}}

\author{Noemi Dreksler}
\affiliation{%
  \institution{Centre for the Governance of AI, Future of Humanity Institute, University of Oxford}
  \city{Oxford}
  \country{UK}
}

\author{Lauren Kahn}
\affiliation{%
 \institution{Perry World House, University of Pennsylvania}
 \city{Philadelphia}
 \state{PA}
 \country{USA}}

\author{Markus Anderljung}
\affiliation{%
  \institution{Centre for the Governance of AI, Future of Humanity Institute, University of Oxford}
  \city{Oxford}
  \country{UK}
}

\author{Allan Dafoe}
\authornote{Author order was randomized using the American Economic Association Author Randomization Tool (Confirmation ID: KiIBo0vkY2Vk).}
\affiliation{%
  \institution{Centre for the Governance of AI, Future of Humanity Institute, University of Oxford}
  \city{Oxford}
  \country{UK}
}

\author{Michael C. Horowitz}
\authornotemark[1]
\affiliation{%
 \institution{Perry World House, University of Pennsylvania}
 \city{Philadelphia}
 \state{PA}
 \country{USA}}

\renewcommand{\shortauthors}{Zewtsloot et al.}

\begin{abstract}
Countries, companies, and universities are increasingly competing over top-tier artificial intelligence (AI) researchers. Where are these researchers likely to immigrate and what affects their immigration decisions? We conducted a survey $(n = 524)$ of the immigration preferences and motivations of researchers that had papers accepted at one of two prestigious AI conferences: the Conference on Neural Information Processing Systems (NeurIPS) and the International Conference on Machine Learning (ICML). We find that the U.S. is the most popular destination for AI researchers, followed by the U.K., Canada, Switzerland, and France. A country's professional opportunities stood out as the most common factor that influences immigration decisions of AI researchers, followed by lifestyle and culture, the political climate, and personal relations. The destination country's immigration policies were important to just under half of the researchers surveyed, while around a quarter noted current immigration difficulties to be a deciding factor. Visa and immigration difficulties were perceived to be a particular impediment to conducting AI research in the U.S., the U.K., and Canada. Implications of the findings for the future of AI talent policies and governance are discussed.
\end{abstract}




\maketitle

\section{Introduction}

\noindent Artificial intelligence (AI) talent is in high demand. Countries, corporations, and universities worldwide are making major investments to cultivate a strong AI talent base \cite{PerraultEtAl2019,Savage2020,VentureScanner}. Though the number of individuals with AI or machine learning (ML) skills is growing, such skills remain highly sought after \cite{LinkedInEmergingJobs,PerraultEtAl2019,Markow2018,ToneyFlagg2020,Gagne2019,WEFJobReport}. Efforts to domestically produce significant AI talent will likely require years to take effect \cite{PennNSCAI}, heightening the need to consider how best to attract global talent.

The immigration preferences of AI researchers have important political, ethical, commercial, and technical implications. These preferences will shape sites of future innovation, the commercial viability of companies, the economic and security prospects of countries, the bargaining power of researchers, and the incentives on employers concerning issues of ethics and governance. As such, better understanding here is relevant to governments, to industry, to those working on the ethics and governance of AI, as well as AI researchers themselves.  

To study AI researchers' immigration preferences, we conducted a survey $(n = 524)$ of AI researchers that had papers accepted at two top-tier AI conferences, the Conference on Neural Information Processing Systems (NeurIPS) and International Conference on Machine Learning (ICML). We focused on understanding where AI talent was planning to move, what drives their immigration decisions, and whether they consider visa and immigration policies as a significant barrier to attracting talent in their country of residence. 

We found that AI researchers are most likely to report considering moving to countries that are well-known for their AI excellence and research hubs. The U.S. was, by far, the most popular potential destination, with 58\% of non-resident AI researchers reporting a 25\% or greater chance of moving there, followed by the U.K. (35\%), Canada (28\%), Switzerland (25\%), and France (16\%). Certain countries --- such as China, France, and Switzerland  ---  showed particular popularity with those AI researchers who had completed their undergraduate degree in the same or nearby countries. By contrast, other countries such as the U.S. and the U.K. were more universally chosen as likely destinations. 

The results show that a variety of personal and professional factors shape AI researchers' immigration decisions. The most commonly chosen factor that drives moving considerations was a country's professional opportunities and environment, selected by 91\% of the respondents as an important immigration factor. Lifestyle and culture, the political climate, and personal relations were all also selected as influencing factors of immigration decisions by a majority of the respondents. Respondents residing in Asia were less likely to report the latter two factors as main considerations in their immigration decisions.

Almost half of the respondents reported that an important factor shaping their thinking was the ease of immigration and/or immigration incentives in the destination country. Similarly, around a quarter of respondents reported that immigration difficulties in their country of residence were an important factor favoring moving. AI researchers residing in the U.S. (69\%), the U.K. (44\%), and Canada (29\%) were most likely to report that visa and immigration problems facing foreign researchers or students in those countries were a significant impediment to conducting high-quality AI research.

Our study offers evidence relevant to how AI researchers choose where to work, how immigration policy affects them, and what this may mean for the future of AI governance. For some countries, such as the U.S. and the U.K., legal immigration hurdles appear to create barriers to attracting AI talent. Other countries, such as China, might face more professional, cultural, and political hurdles in making immigration attractive to AI researchers. Before we turn to the survey results in detail, we contextualize our findings by discussing the relationship between immigration and innovation, the AI talent policy landscape, and the limited existing research on AI researchers' immigration motivations and intentions.

\section{Background}

\subsection{Immigration and Innovation}

\noindent There is a steadily growing literature on the association between immigration, entrepreneurship, and innovation. The empirical evidence from North America and Europe points to a positive relationship between immigration and innovation measures such as patent applications; this research also highlights the large share of new businesses started by immigrants \cite{Kauffman2017,Kerr2019,KerrOzden2017,OzgenNijkamp}. Immigrants appear to have --- on balance --- a positive effect on the output of industry and contribute strongly to academia \cite{Stephan2010}. \citeauthor{HuntGauthier-Loiselle} \cite{HuntGauthier-Loiselle}, for example, found for U.S. state panel data from 1940 to 2000 that there was a 9-18\% increase in patents per capita for each percentage point increase in the share of immigrant college graduates in the total population. Similarly, for a panel of 20 European countries, \citeauthor{BosettiCattaneo} \cite{BosettiCattaneo} found that skilled worker migration increased knowledge creation in both the public and private sphere based on patenting and bibliometric data from 1995 to 2008.

In turn, innovation is generally seen as essential for economic growth  \cite{Cameron1996,Fagerberg2006,Verspagen2006}. Taken together, the strategic and economic importance of AI capabilities, the limited supply yet growing demand of AI-related skills, and the positive relationship between immigration and innovation, point to it being in countries' and businesses' interest to understand the AI talent pool and what drives their immigration decisions.

\subsection{Governmental AI Talent Strategies and Policies}

\noindent The importance of AI and the associated need to foster and attract AI talent is reflected in the strategic objectives and changing immigration policies of countries worldwide, including the U.K., Australia, Canada, and France \cite{HuangArnold2020}. As AI has come to be viewed as strategically important, many countries have developed AI strategies and high-level AI principles \cite{CIFAR2018,Jobinetal2019}. In step with such strategic aims, there has been an accompanying focus on building domestic talent, improving AI-related and digital literacy skills, and attracting talent through immigration.

For example, one of the four core objectives of Canada's 2017 Pan-Canadian Artificial Intelligence Strategy is to ``[a]ttract and retain world-class AI researchers by increasing the number of outstanding AI researchers and skilled graduates in Canada,'' and has been successful in recruiting well-known international AI researchers to Canadian universities and research institutes \cite{Pan-CanadianAIStrategy,topAITalentCanada,CanadaRecruitment2020,Semuels2019}. With the 2017 release of the ``New Generation Artificial Intelligence Development Plan'' by China's State Council, China announced its intention to become the world leader in AI by 2030.\footnote{For discussion of these policies, see \citeauthor{RobertsCowlsMorley} \cite{RobertsCowlsMorley}.} In service of this aim, China has instituted numerous state- and party-sponsored initiatives to recruit talent both within and outside of its borders, ``to bolster its strategic civilian and military goals'' \cite{ChinaTalentTracker}. 

Similarly, in the last few years, European countries have also implemented immigration policies to facilitate those working in technical fields such as AI to stay and work in the country \cite{HuangArnold2020}. For example, in 2016 and 2017, France created the ``French Tech Ticket'' and the ``French Tech Visa'' to encourage technology startups \cite{TeamYS}. While in the U.K., the 2020 start of a ``point-based immigration system'' is hoped to attract more highly skilled workers \cite{Gov.uk}.

There are decades worth of empirical data pointing to the centrality of immigrants to the U.S. technology sector \cite{BrownEarleKim,Saxenian1999,WadhwaSaxenian} and this trend appears to hold for the AI sector: half of the U.S.' AI degree graduates and two-thirds of graduate students in AI-related graduate programs were born elsewhere \cite{Zwetsloot2019,AIDataVisualization}, while 66\% of the U.S.'s 50 ``most promising'' AI startups have at least one first-generation immigrant founder \cite{HuangArnoldZwetsloot2020}.

And yet, in contrast with other countries looking to enhance their AI capabilities, the U.S. has not reformed its immigration system to make it easier for highly-skilled AI and STEM professionals to come and stay in the country \cite{HuangArnold2020}. A U.S. executive order suspending the issuing of H-1B visas in June 2020 amidst the COVID-19 pandemic is likely to have only made it harder \cite{Hao2020}. AI experts fear that a failure to implement reforms will undermine U.S. AI efforts \cite{AAAINITRD,Etzioni2019}. On this point, the U.S. National Security Commission on Artificial Intelligence's (NSCAI) most recent Interim Report emphasizes that reforming U.S. immigration and visa laws play a critical role in ``ensuring U.S. national security and furthering the interests of democratic nations around emerging technologies'' \cite{NSCAI}. Industry experts, too, have repeatedly highlighted the curtailing effect such lacking immigration policies have on the U.S.' collective AI productivity \cite{Arnold2019DefenseOne}. It is, perhaps, unsurprising then that U.S. residents were more likely than residents of other countries to report immigration difficulties as a problem for conducting high-quality AI research in their country of residence. 

\subsection{Existing Research on AI Researchers' Immigration Preferences}

\noindent While governments are taking actions to build and attract AI talent, there is limited research dedicated to surveying the individuals of interest--for example, AI researchers themselves. Studying AI researchers' moving preferences but also how such policies may be affecting them is important in understanding the human side of immigration decisions. There is a growing academic literature that surveys the AI researcher community. Still, most of these surveys have focused on eliciting forecasts on AI, how they define AI, or the perceived impact of AI \cite{SandbergBostrom2011,BaumGoertzl2,MullerBostrom,GraceEtAl,zhang2020us,walsh2018expert,gruetzemacher2019forecasting,krafft2020defining,anderson2018artificial}. 

Research on the AI talent pool in terms of their career and immigration preferences, has seen an increasing interest only in recent years. Such research is particularly salient amongst private companies, think tanks, and research and governmental institutions \cite{EuroCommission,Roca2019}. For example, the Center for Security and Emerging Technology (CSET) has an entire report series on global AI talent \cite{GehlhausMutis2021,HuangArnold2020,2Zwetsloot2019}. 

Since we want to better understand AI talent’s immigration decisions, of particular note for our study is a recent survey of PhD students from top-ranking U.S. AI universities, conducted by \citeauthor{Aiken2020} \cite{Aiken2020}. It found that the quality of the education and future job opportunities were the two most commonly chosen reasons for choosing a PhD program in the U.S. In addition, growth opportunities, colleagues, research interests, technical challenges, and location were amongst the most ``extremely important'' rated factors for job attractiveness. The same research team found that the AI PhDs who chose to leave the U.S. were likely to cite immigration-related concerns (23\%) and the U.S. immigration system (33\%) as highly relevant to their decision to leave \cite{AikenDunham2020}. Even more commonly chosen reasons to leave the U.S. were job opportunities (58\%) and family and friends (57\%) abroad. 

Finally, a number of studies have been conducted that explore the immigration pathways of AI talent inferred from publicly available data on the researchers that had papers accepted at prestigious AI and ML conferences. Such data cannot give us direct insights into AI researchers' immigration preferences. However, it can be seen as indirect evidence of the immigration preferences that they were able to act on successfully. The sampling frames often coincide with that of our study: authors with papers accepted at NeurIPS and ICML. Analysts usually study these AI researchers' institutional affiliations, where they have moved over the course of their education and career, and measures of their research success. Notable examples include JF Gagne's 2019 AI Talent Report \cite{Gagne2019} and Macro Polo’s Global Talent Tracker \cite{GlobalTalentTracker}. These analyses point to a global and highly mobile talent pool that is likely to migrate to other countries over the course of their academic and professional career. The U.S., the U.K., China, Canada, Germany, Switzerland, and France stand out as attracting the most top-tier AI talent and ranking highest in AI research output at the investigated conferences \cite{Chuvpilo2018,Chuvpilo2019,2Chuvpilo2020}. The U.S. is found to have a resounding lead in AI research in terms of the number of top AI researchers trained there and who immigrated there for work. Countries such as China and India contribute a large fraction of this immigrant AI talent.

The current study fills a crucial gap in the literature by combining the direct surveying of AI researchers' immigration preferences also seen in studies by \citeauthor{AikenDunham2020} \cite{Aiken2020,AikenDunham2020}, with a broader sample not restricted to U.S.-trained researchers. Instead, our study is drawn from the authors at two prestigious AI conferences --- a sampling frame used only by studies to understand AI talent's immigration patterns by collecting publicly available data, rather than surveying the researchers themselves. This means our study offers new insights into AI researchers' immigration decisions through the eyes of this highly sought-after sample of top-tier AI talent themselves.

\section{Methods}

\begin{figure*}[h]
\centering
\includegraphics[width=0.7\textwidth]{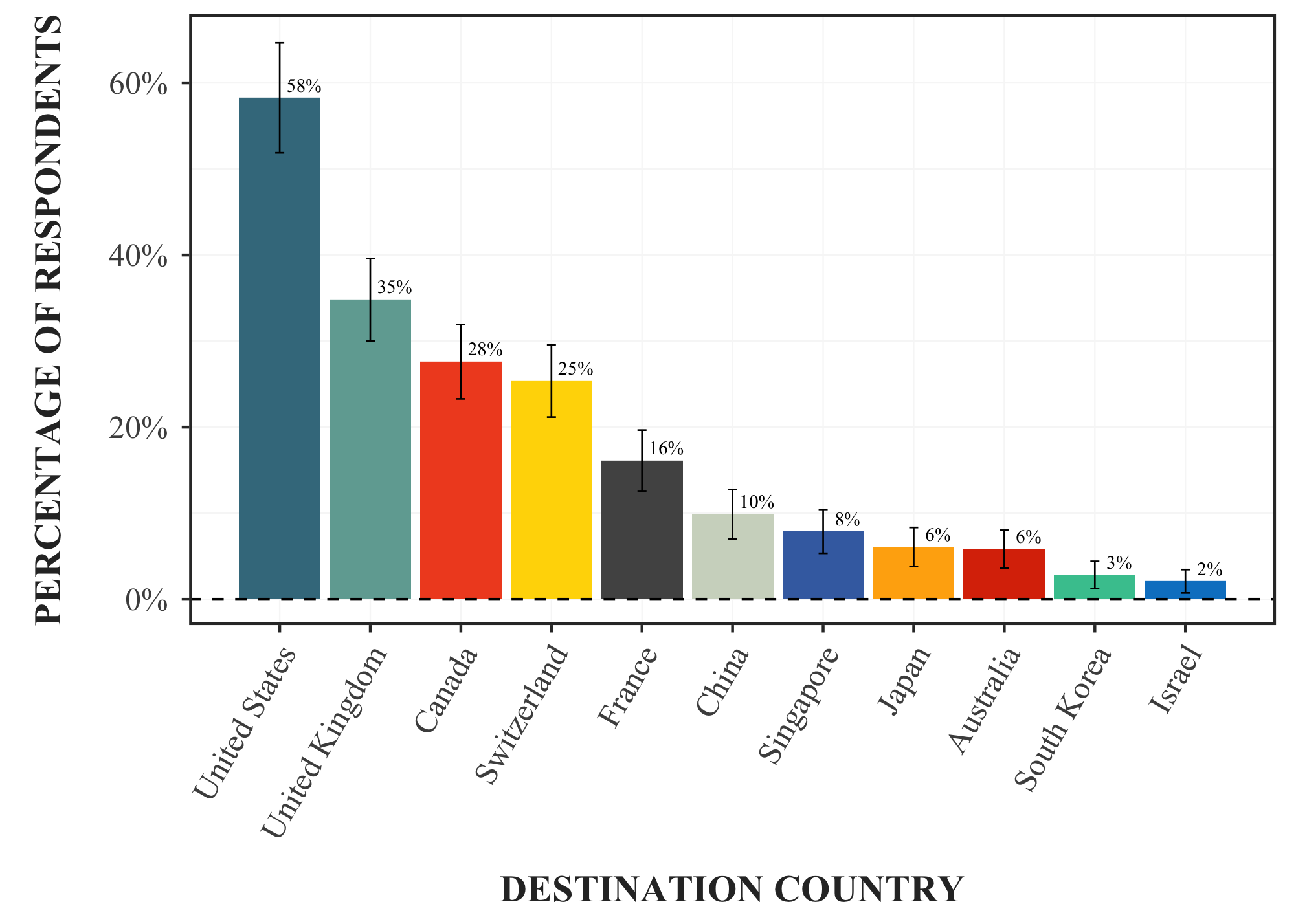}
\caption{Countries AI researchers are most likely to move to | The graph shows the percentage of AI researchers who reported a greater than 25\% chance of moving to a country for work or study within the next three years. Respondents were presented with a list of countries to choose from (see Appendix for a full list). Each respondent's country of residence was omitted from the list. There was also a ``none of these'' and an ``other'' option in which respondents could write a different country they had a greater than 25\% chance of moving to. Error bars represent 95\% confidence intervals.}
\label{fig1}
\end{figure*}

\noindent AI researchers were surveyed between September 16 and October 13, 2019 and answered questions about their attitudes towards AI, as well as AI policy and governance topics. The results presented in this paper are based on the immigration-related questions that formed part of the larger survey \cite{ZhangAnderljungetal}. The full survey text of the questions presented in this paper can be found in the Appendix. The survey took a median of 17 minutes to complete.

Respondents were selected based on having papers accepted at two top AI research conferences, using the same sampling frame as \citeauthor{GraceEtAl} \cite{GraceEtAl}. One group of respondents was drawn from researchers who published papers at the 2018 NeurIPS conference. Another group of respondents was drawn from those who published articles at the 2018 ICML conference. An additional group of respondents was drawn from samples who participated in a 2016 expert survey on AI \cite{GraceEtAl} and had papers accepted at NeurIPS and ICML in 2015. 

A total of 3,030 researchers were contacted, and 17\% of the researchers $(n = 524)$ completed the survey. Respondents worked or studied in 27 different countries from around the world, although the majority were residents in North America or Western Europe (residence data was missing for 72 respondents). Countries with the most respondents resident there were: the U.S. $(n = 209)$, the U.K. $(n = 54)$, France $(n = 25)$, Canada $(n = 22)$, Germany $(n = 22)$, Switzerland $(n = 22)$, and China $(n = 21)$. We used publicly available information to collect demographic data about the respondents (e.g., country of undergraduate degree, type of workplace, citation count measures, etc.). The information on the analysis of the representativeness of the sample can be found in the Appendix.

The analysis was pre-registered on the Open Science Framework.\footnote{Project accessible here: \url{https://osf.io/fqz82/}} ``Don’t know'' or missing responses were re-coded to the weighted overall mean, unconditional on treatment conditions. Heteroscedasticity-consistent standard errors were used to calculate the 95\% confidence level margins of error shown on the graphs.

\begin{figure*}[h]
\centering
\includegraphics[width=0.8\textwidth]{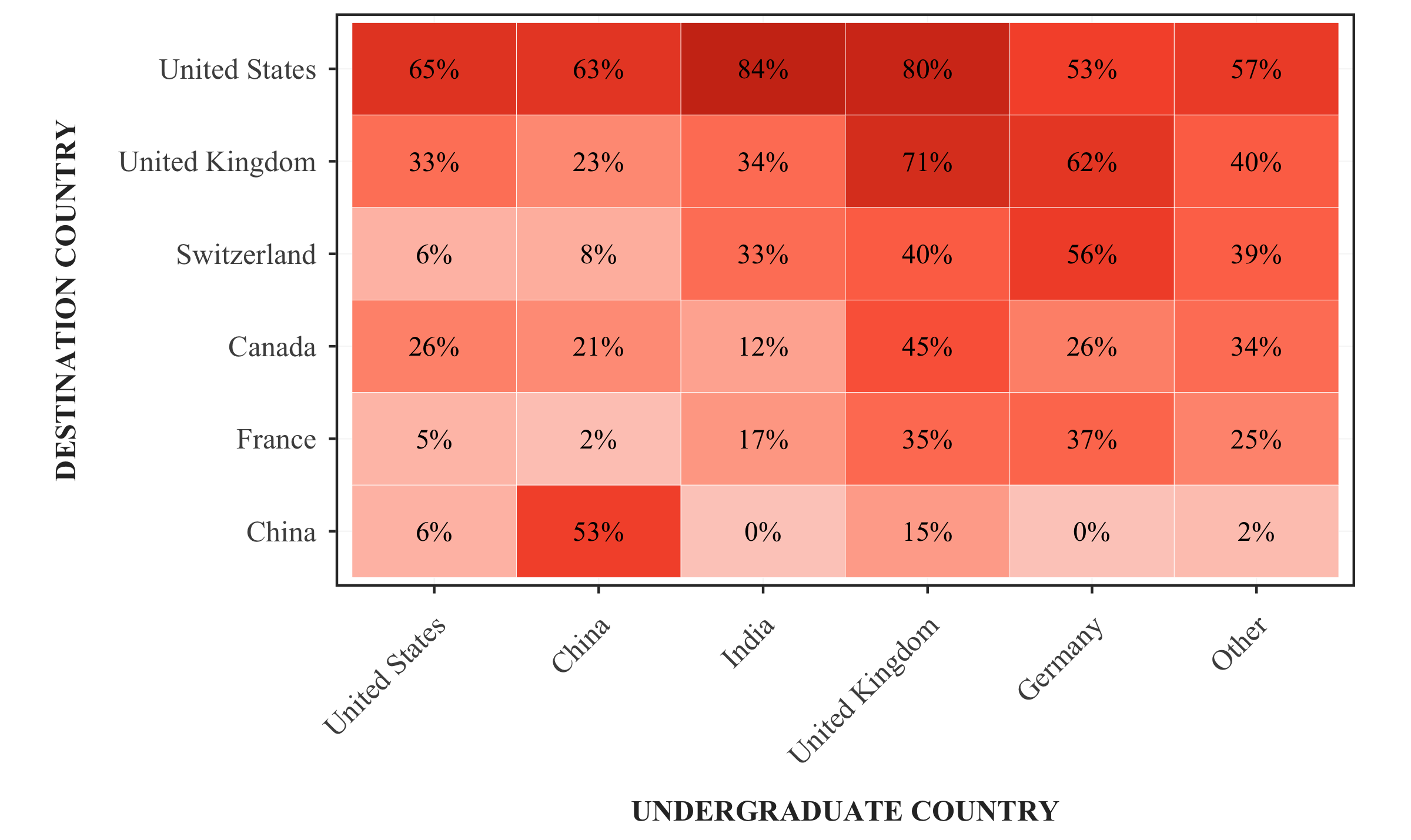}
\caption{Likely moving destinations by country of undergraduate degree | Respondents were asked where they had a greater than 25\% chance of moving to for work or study within the next three years. They were presented with a list of countries to choose from as well as ``none of these'' and ``other'' options (see Appendix for the full list). Respondents' country of residence was omitted from the list. This figure shows the percentage of AI researchers who selected the six most popular of the possible destination countries, broken down by where respondents completed their undergraduate degree.}
\label{fig2}
\end{figure*}

\section{Results}

\noindent The three questions we asked respondents that related to immigration issues aimed to answer three main queries: What countries are AI researchers most likely to move to? What factors affect these moving decisions? Do AI researchers judge visa and immigration policies as a serious problem for conducting high-quality AI research in their country of residence? Below we present the results for each of these questions in turn.

\subsection{AI Researchers’ Likelihood of Migrating}

\noindent In order to understand the appeal of different countries to AI researchers and how likely it was AI researchers might move there, respondents were asked: ``Of the following countries, which would you have a greater than 25\% likelihood of moving to for work or study in the next three years?'' The list of ten countries presented omitted the respondent’s country of residence, and there was a ``none of these'' response option and an ``other'' text box where respondents could enter a different country. The countries were chosen to reflect those countries which have been shown to be top destinations for AI researchers \cite{Gagne2019,PerraultEtAl2019,2Zwetsloot2019}.

In Figure \ref{fig1} the percentage of participants that reported a greater than 25\% chance of moving to a country they are currently not resident in are shown. The U.S. is by far the most popular destination, with a majority of non-U.S. residents (58\%) reporting they would consider moving there in the next three years. Indeed, taken together with the 209 respondents who already lived in the U.S., only 22\% of our sample did not already live in the U.S. nor were considering moving there in the next three years. The U.K. (35\%), Canada (28\%), Switzerland (25\%), and France (16\%) were also popular destinations for AI researchers. Many of these countries have already begun optimizing their immigration systems to increase the amount of AI and tech talent that can flow into the country \cite{Arnold2019}. These results generally replicate past research conducted on AI PhD students, that similarly found that the U.S., Canada, the U.K., Germany, and Switzerland were amongst the most likely destinations to move to after graduation \cite{AikenDunham2020}.

It may be surprising that a country such as China, which performs strongly in AI research and invests heavily in AI technology \cite{Hao2019}, is not seen as a more attractive country for top global AI talent. However, if we look more closely at the data, we can see that AI researchers from different backgrounds may be differentially attracted to specific destinations.

Figure \ref{fig2} shows the most popular destination countries by the country where a respondent completed their undergraduate degree. The country where someone completes their undergraduate degree is sometimes taken as a proxy for citizenship where such data is not available.\footnote{Students often complete their undergraduate degree in the country where they were born. While this can misclassify certain respondents if one were to take the country a respondent completed their undergraduate degree as a proxy for citizenship, the misclassification rate generally remains low. \citeauthor{Zwetsloot2019} \cite{Zwetsloot2019} estimated it could be up to 20\% for their sample of AI talent. Nevertheless, care has to be taken when one makes such an inference of citizenship or nationality from an undergraduate degree country measure. Thus, such conclusions should be held with an adequate amount of uncertainty when couched in this way. We note that this is also the case for our later analysis in this vein.} The figure shows that some countries remain invariably popular across the sample (e.g., the U.S.' appeal as a destination varies between 53\% and 84\%). Other countries clearly show stronger regional or country-specific appeal. Respondents who completed their undergraduate degree in China are more likely to consider moving to China in the future (53\%). The U.K. appears to be a particularly popular destination for individuals who completed their undergraduate degree in Germany and the U.K. We can see a similar regional effect for France and Switzerland: they appear to be most appealing to respondents with undergraduate degrees from other countries in Europe, like the U.K. and Germany, while being the least appealing to those with undergraduate degrees from the U.S. or China.

\begin{figure*}[h]
\centering
\includegraphics[width=0.7\textwidth]{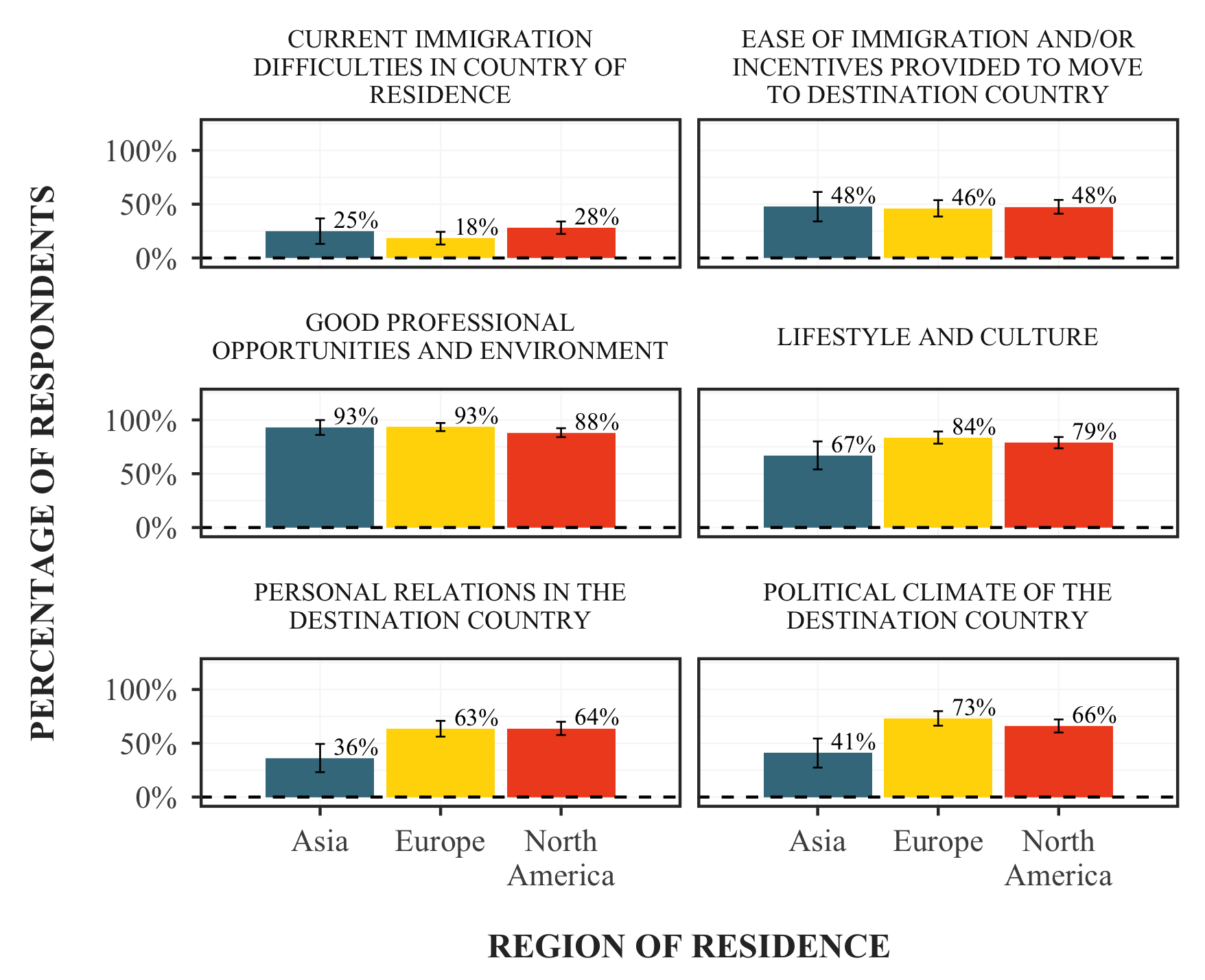}
\caption{Factors affecting AI researchers’ immigration decisions by region of residence | Respondents were asked what factors were important in their consideration of moving to work or study in a country that they don’t currently work or study in full-time. Respondents were presented with the six options shown in the graph alongside an ``other'' and a ``none of the above'' option. The figure shows the percentage of AI researchers who reported that these six factors were important, broken down by their current region of residence (Asia, Europe, and North America). China, South Korea, Japan, India, and Singapore were coded as Asia. Russia was coded as Europe alongside European Union countries, which at the time of the survey still included the U.K. The U.S. and Canada were coded as North America. Error bars represent 95\% confidence intervals. 
}
\label{fig3}
\end{figure*}

A country's ability to attract talent is essential, but retaining it is just as important. When we examine the percentage of respondents who chose ``none of these'' to the above question (i.e., expressing no intention to move) it was here too that the U.S. appeared to be the most attractive destination for AI researchers. Out of the AI researchers resident in the U.S., 37\% reported no moving intentions with a high likelihood over the next three years. Germany (23\%) and France (21\%) also appeared to have high rates of AI researchers with no intention to move. In China, Switzerland, Canada, and the U.K, between 14 and 17\% of current residents reported no intention to move to another country. It should be noted that even the highest rate in the U.S. of 37\% still speaks to a highly mobile workforce, with at least around two thirds or more of top AI researchers in each country being prepared to consider a move to where working and living conditions best coincide with their aspirations and desires with at least a 25\% chance likelihood.

\subsection{AI Researchers’ Immigration Decisions}

\begin{figure}[h]
\centering
\includegraphics[width=\linewidth]{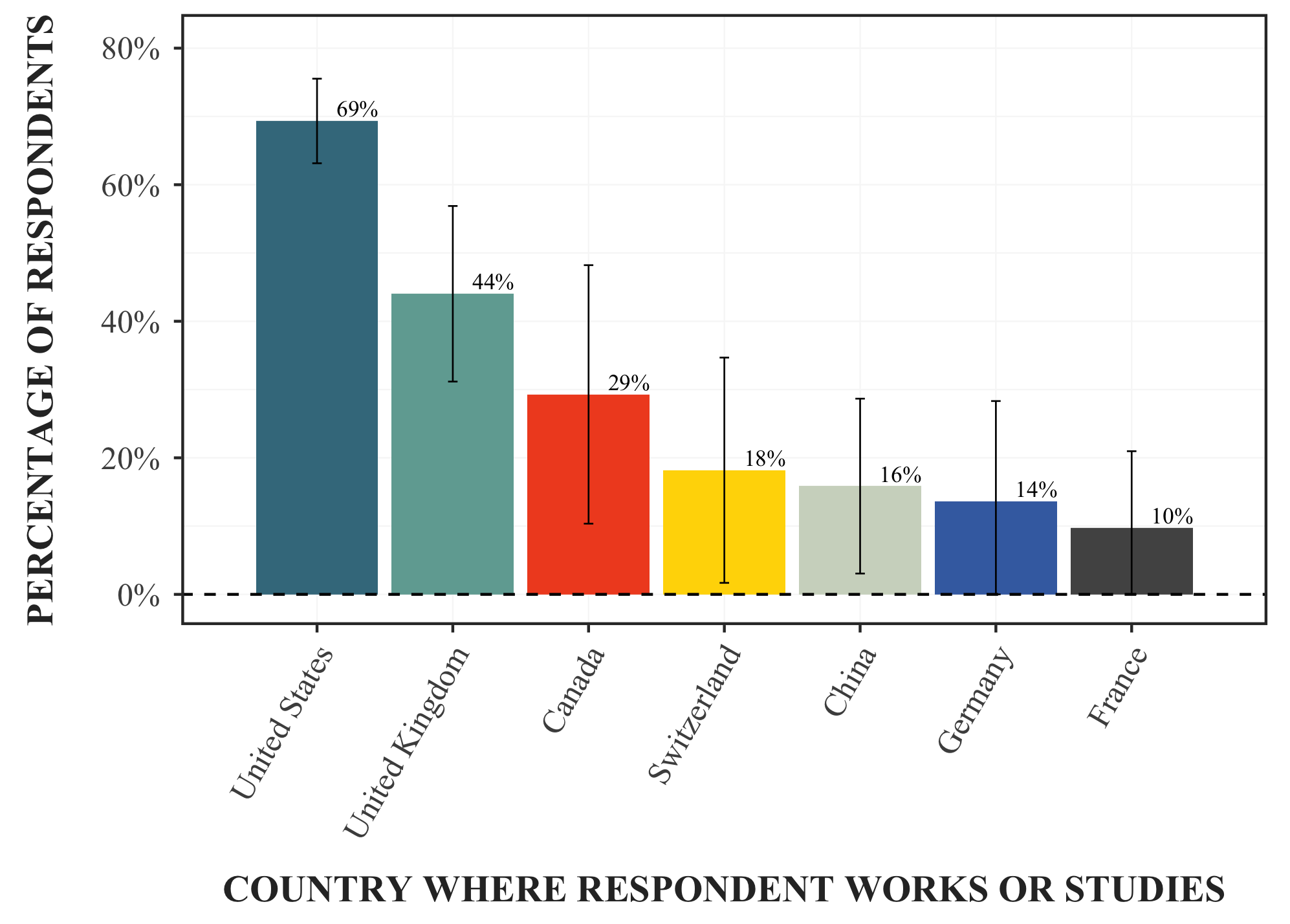}
\caption{Where visa and immigration issues are considered a serious problem for conducting AI research | Respondents were asked what they considered to be serious problems for conducting high-quality AI research in their country of residence, with a list of option that included seven issues (see Appendix). The figure presents the by-country breakdowns for one of these issues: the percentage of AI researchers living in each country who feel that visa and immigration issues are a serious problem for conducting high-quality AI research in their country of residence. Error bars represent 95\% confidence intervals. 
}
\label{fig4}
\end{figure}

\noindent Studies that examine the movement or geography of AI talent across the globe \cite{Gagne2019,GlobalTalentTracker} cannot directly reveal what may be influencing, aiding, and impeding AI researchers’ intentions and immigration decisions. To probe what factors may be important to AI researchers in their immigration decisions we asked respondents: ``When considering moving to work or study in a country that you don’t currently work or study in full-time, what factors are important in your consideration?'' Respondents were able to select as many options as they wanted from the list of six factors shown in the Appendix, alongside an ``other'' option where they could fill in their own text. These options were chosen since the literature indicates that these are critical drivers of high-skill immigration decisions \cite{HanAppelbaum,MusumbaJinMjelde,2Zwetsloot2019}.

The most common factor in immigration decisions that AI researchers reported was whether a country provides good professional opportunities and environment, with 91\% of respondents reporting this affects their moving considerations. This is mirrored by previous research on US-trained AI PhDs, which found that job-related factors, such as growth opportunities, colleagues, research interests, and technical challenges were rated highly in terms of contributing to job attractiveness \cite{Aiken2020}. Lifestyle and culture (79\%), the political climate of the destination country (66\%), and having personal relations in the destination country (60\%) were all also chosen by a majority of the respondents. Just under half of the respondents (47\%) cited that the ease of immigration or incentives provided by the destination country influence their immigration decisions. Finally, around one quarter (24\%) of respondents noted that current immigration difficulties in their country of residence were important in their moving considerations. 

A quarter of a valuable and highly skilled workforce in high demand might appear comparatively low to the other factors but could still be a critical loss for a country's AI talent pool. However, the phrasing of the question does not make it possible to determine whether this comparatively lower percentage is because respondents are not encountering difficulties (perhaps because they are citizens in their country of residence or have secured the required long-term visas without issue), or whether this simply is not a factor that AI researchers feel affects their immigration considerations. Nevertheless, previous research does speak to the fact that this highly skilled workforce is experiencing immigration difficulties in securing visas for themselves or family members and partners, at least in the U.S. \cite{AikenDunham2020}.

Another way of approaching this data is by looking at what potential non-citizens respond to this item by taking the country of undergraduate degree as a proxy for citizenship. As noted previously, this is a commonly used proxy, but should be treated with an adequate level of caution in terms of the certainty of conclusions drawn from such analysis due to the potential for misclassification. Taking this into account, we can then consider proxy-non-citizens respondents to be those whose undergraduate country does not correspond with their current country of residence. This analysis reveals that proxy non-citizen respondents reported current immigration difficulties to be a consideration to a different extent by country of residence. In the U.S., 40\% of proxy non-citizens reported current immigration difficulties as a factor in their immigration decisions. The percentage is slightly lower in Canada (31\%) and the U.K. (21\%), and far lower in Switzerland (14\%). Sample sizes were exceedingly small for this subgroup of respondents in other countries, so they could not be representatively analyzed. 

Finally, it is worth considering whether the factors influencing AI researchers' moving decisions may differ depending on where they work and live. Figure \ref{fig3} shows what percentage of respondents chose each factor as important in their moving considerations based on their region of residence. For some factors, the weight placed on certain factors appears to be uniform across regions: Good professional opportunities and environment is the consistently most chosen factor across regions and just under half of the respondents in all regions consider ease of immigration and incentives provided in the destination country as an important moving factor. The political climate and having personal relations in the destination country appears to be chosen less often in Asia as an important factor than in Europe, with the same trend being seen to a lesser degree for lifestyle and culture. Respondents resident in Europe reported current immigration difficulties least often as a moving factor compared to residents of Asia and North America; however, this should not be interpreted as a significant difference due to the larger error bars, likely afforded by the smaller sample sizes of these subgroups.

\subsection{AI Researchers' Perceptions of Immigration Policy}

\noindent It is clear that there is one key hurdle for attracting and retaining high-skilled talent even if a country is attractive to AI researchers in and of itself: the immigration and visa system of a country has to make it legally possible for AI researchers and engineers to be able to enter the country, work, and live there long-term, along with any partners or family they may have. 

Respondents were asked about a number of issues and whether they felt they created a serious problem for conducting high-quality AI research in their country of residence. The full list of issues can be found in the Appendix. In Figure \ref{fig4} we present the percentage of respondents who reported that ``visa and immigration problems facing foreign researchers or students who want to work or study'' in the country are a serious problem for conducting high-quality AI research. As can be seen in Figure \ref{fig4}, 69\% of U.S. residents reported this as a serious issue, far more than residents from other countries.

This particular sample of U.S. resident AI researchers is not alone in holding this belief: U.S. companies, researchers, and public intellectuals have criticized the overly restrictive immigration policies of the U.S. \cite{Arnold2019,HuangArnold2020}. They have highlighted that such policies are deleterious to the U.S. AI sector, in particular when other countries have improved and better-designed immigration policies that make it easier to attract and retain talent. Indeed, in a recent study of U.S.-trained AI PhDs, 60\% of non-citizens reported having had significant problems with the U.S. immigration system \cite{AikenDunham2020}. This stands in stark contrast to the 12\% of non-citizen AI PhDs not resident in the U.S. who reported such issues with the immigration system in their country of residence.

In our study, a notable percentage of AI researchers in the U.K. (44\%) and Canada (29\%) reported visa and immigration issues as a serious problem for AI research. However, only between 10\% and 18\% of respondents in Switzerland, China, Germany, and France felt that visa and immigration facing students and workers were a serious problem for high-quality AI research in their country of residence. The lessons from such findings are not immediately clear. Perhaps, such countries where workers report visa and immigration policies as less of an issue simply do not have further demand for AI talent. However, this would be at odds with what European governments perceive their own need for AI talent to be \cite{Anderson2020}. Nevertheless, what does seem clear is that it is imperative for countries such as the U.S., the U.K., and to a lesser degree Canada, to remain vigilant and proactive in how their immigration policies may be standing in the way of them successfully attracting and retaining AI researchers.

\section{Conclusion}

\noindent Decisions by AI researchers about where to work could have significant consequences for AI governance. The way countries design immigration and other policies to attract AI talent could influence everything from government policies surrounding AI ethics to how countries approach broader questions about the appropriate and safe use of AI. Understanding the preferences and motivations of AI researchers is also a critical step for policy-makers and businesses in strengthening their AI ecosystems and talent bases. Our findings highlight where the potential pressure points are for countries and what could make them more welcoming and their environment more attractive for AI talent. Beyond the relevance to ethics, governance, and business, our findings can be informative to AI researchers themselves. The study offers an insight into how AI researchers might be affected by the immigration policies of different countries. Knowing where visa and immigration issues are considered a problem for conducting high-quality AI research can help researchers decide where to take their next career steps. Uprooting one’s life, whether alone or with a partner or family, is never a small decision, even for a workforce as mobile as AI researchers.

Visa and immigration policies that fail to be tailored to attracting highly skilled workers successfully, as seems to be the case for the U.S., do appear to be experienced as a serious impediment for AI research and are a likely net loss for countries in terms of their ability to attract and retain AI talent. Even countries that have taken more targeted steps to ease high-skilled immigration, such as the U.K. and Canada \cite{HuangArnold2020}, still have substantial portions of AI researchers reporting that visa and immigration issues are problems for conducting high-quality AI research in their country of residence.

Other countries such as China may be facing not a legal barrier to attracting talent but rather the challenge of increasing the overall appeal of the professional and cultural opportunities and environment they can offer AI researchers. The political climate could also be a contributing factor. The key to being an attractive destination for AI talent may not lie simply in attaining research excellence and having available funding either: In a 2012 survey of 2,314 Nature readers --- consisting mainly of U.S. and European STEM academics --- more than 60\% of respondents reported that the best biological and physical science would be happening in China by 2020, while only 8\% reported they would relocate there \cite{VanNoorden2012}. The U.S., Canada, Europe, and Australia were seen as more attractive, as they currently appear to be for most AI talent.

Countries seeking to attract AI talent need to understand what drives the immigration decisions of AI researchers. Furthermore, an important ethical imperative is to understand AI researchers' immigration experiences and motivations on a human level. AI researchers' immigration attitudes and actions are likely to have wide-reaching reverberations: they have the power to shape the future of AI governance and ethics, as well as global immigration policies for highly skilled workers. Future research could address the limitations of this single cross-sectional study by extending the sampling frame beyond NeurIPS and ICML. Furthermore, researchers could ask more fine-grained questions about what has driven immigration decisions or caused problems in specific instances of trying to immigrate to particular countries. Finally, researchers could survey a panel of AI researchers across time to better understand how immigration preferences and motivations shift at different career junctures. 

\begin{acks}
This project was supported by the Ethics and Governance of Artificial Intelligence Fund, and Michael Horowitz’s work on this project was supported in part by U.S. government grant FA9550-18-1-0194. The authors would like to thank Catherine Aiken, Zachary Arnold, Tessa Baker, James Dunham, Melissa Flagg, Charlie Giattino, Roxanne Heston, Igor Mikelic-Torreira, Dewey Murdick, and Helen Toner for feedback on the AI expert survey and the analysis. We are also grateful for research assistance and editorial support from Emmie Hine, Will Marks, Kwan Ye Ng, and Sacha Zimmerman.
\end{acks}


\begin{thebibliography}{67}


\ifx \showCODEN    \undefined \def \showCODEN     #1{\unskip}     \fi
\ifx \showDOI      \undefined \def \showDOI       #1{#1}\fi
\ifx \showISBNx    \undefined \def \showISBNx     #1{\unskip}     \fi
\ifx \showISBNxiii \undefined \def \showISBNxiii  #1{\unskip}     \fi
\ifx \showISSN     \undefined \def \showISSN      #1{\unskip}     \fi
\ifx \showLCCN     \undefined \def \showLCCN      #1{\unskip}     \fi
\ifx \shownote     \undefined \def \shownote      #1{#1}          \fi
\ifx \showarticletitle \undefined \def \showarticletitle #1{#1}   \fi
\ifx \showURL      \undefined \def \showURL       {\relax}        \fi
\providecommand\bibfield[2]{#2}
\providecommand\bibinfo[2]{#2}
\providecommand\natexlab[1]{#1}
\providecommand\showeprint[2][]{arXiv:#2}

\bibitem[\protect\citeauthoryear{Aiken, Dunham, and Zwetsloot}{Aiken
  et~al\mbox{.}}{2020a}]%
        {Aiken2020}
\bibfield{author}{\bibinfo{person}{Catherine Aiken}, \bibinfo{person}{James
  Dunham}, {and} \bibinfo{person}{Remco Zwetsloot}.}
  \bibinfo{year}{2020}\natexlab{a}.
\newblock \bibinfo{booktitle}{\emph{{Career Preferences of AI Talent}}}.
\newblock \bibinfo{type}{Data Brief}. \bibinfo{institution}{Center for Security
  and Emerging Technology}.
\newblock
\urldef\tempurl%
\url{https://cset.georgetown.edu/research/career-preferences-of-ai-talent/}
\showURL{%
\tempurl}


\bibitem[\protect\citeauthoryear{Aiken, Dunham, and Zwetsloot}{Aiken
  et~al\mbox{.}}{2020b}]%
        {AikenDunham2020}
\bibfield{author}{\bibinfo{person}{Catherine Aiken}, \bibinfo{person}{James
  Dunham}, {and} \bibinfo{person}{Remco Zwetsloot}.}
  \bibinfo{year}{2020}\natexlab{b}.
\newblock \bibinfo{booktitle}{\emph{{Immigration Pathways and Plans of AI
  Talent: Findings from a CSET Survey of Artificial Intelligence PhDs from U.S.
  Universities}}}.
\newblock \bibinfo{type}{Data Brief}. \bibinfo{institution}{Center for Security
  and Emerging Technology}.
\newblock
\urldef\tempurl%
\url{https://cset.georgetown.edu/research/immigration-pathways-and-plans-of-ai-talent/}
\showURL{%
\tempurl}


\bibitem[\protect\citeauthoryear{Anderson, Rainie, and Luchsinger}{Anderson
  et~al\mbox{.}}{2018}]%
        {anderson2018artificial}
\bibfield{author}{\bibinfo{person}{Janna Anderson}, \bibinfo{person}{Lee
  Rainie}, {and} \bibinfo{person}{Alex Luchsinger}.}
  \bibinfo{year}{2018}\natexlab{}.
\newblock \bibinfo{booktitle}{\emph{Artificial Intelligence and the Future of
  Humans}}.
\newblock \bibinfo{type}{{T}echnical {R}eport}. \bibinfo{institution}{Pew
  Research Center}.
\newblock
\urldef\tempurl%
\url{https://perma.cc/4QG4-VM4F}
\showURL{%
\tempurl}


\bibitem[\protect\citeauthoryear{Anderson, Viry, and Wolff}{Anderson
  et~al\mbox{.}}{2020}]%
        {Anderson2020}
\bibfield{author}{\bibinfo{person}{Julia Anderson}, \bibinfo{person}{Paco
  Viry}, {and} \bibinfo{person}{Guntram~B. Wolff}.}
  \bibinfo{year}{2020}\natexlab{}.
\newblock \showarticletitle{{Europe has an artificial-intelligence skills
  shortage}}.
\newblock \bibinfo{journal}{\emph{Bruegel}} (\bibinfo{year}{2020}).
\newblock
\urldef\tempurl%
\url{https://www.bruegel.org/2020/08/europe-has-an-artificial-intelligence-skills-shortage/}
\showURL{%
\tempurl}
\newblock
\shownote{Accessed January 21, 2021.}


\bibitem[\protect\citeauthoryear{Arnold}{Arnold}{2019}]%
        {Arnold2019DefenseOne}
\bibfield{author}{\bibinfo{person}{Zachary Arnold}.}
  \bibinfo{year}{2019}\natexlab{}.
\newblock \showarticletitle{{Misguided Immigration Policies Are Endangering
  America's AI Edge}}.
\newblock \bibinfo{journal}{\emph{DefenseOne}} (\bibinfo{year}{2019}).
\newblock
\urldef\tempurl%
\url{https://www.defenseone.com/ideas/2019/11/misguided-immigration-policies-are-endangering-americas-ai-edge/161366/}
\showURL{%
\tempurl}
\newblock
\shownote{Accessed January 21, 2021.}


\bibitem[\protect\citeauthoryear{Arnold, Heston, Zwetsloot, and Huang}{Arnold
  et~al\mbox{.}}{2019}]%
        {Arnold2019}
\bibfield{author}{\bibinfo{person}{Zachary Arnold}, \bibinfo{person}{Roxanne
  Heston}, \bibinfo{person}{Remco Zwetsloot}, {and} \bibinfo{person}{Tina
  Huang}.} \bibinfo{year}{2019}\natexlab{}.
\newblock \bibinfo{booktitle}{\emph{{Immigration Policy and the U.S. AI
  Sector}}}.
\newblock \bibinfo{type}{{T}echnical {R}eport}. \bibinfo{institution}{Center
  for Security and Emerging Technology}.
\newblock
\urldef\tempurl%
\url{https://cset.georgetown.edu/research/immigration-policy-and-the-u-s-ai-sector/}
\showURL{%
\tempurl}


\bibitem[\protect\citeauthoryear{{Association for the Advancement of Artificial
  Intelligence}}{{Association for the Advancement of Artificial
  Intelligence}}{2018}]%
        {AAAINITRD}
\bibfield{author}{\bibinfo{person}{{Association for the Advancement of
  Artificial Intelligence}}.} \bibinfo{year}{2018}\natexlab{}.
\newblock \bibinfo{title}{{AAAI Response to NITRD RFI: National Artificial
  Intelligence Research and Development Strategic Plan}}.
\newblock
\newblock
\urldef\tempurl%
\url{https://www.nitrd.gov/nitrdgroups/index.php?title=AI-RFI-Responses-2018}
\showURL{%
\tempurl}
\newblock
\shownote{Accessed January 21, 2021.}


\bibitem[\protect\citeauthoryear{Baum, Goertzel, and Goertzel}{Baum
  et~al\mbox{.}}{2011}]%
        {BaumGoertzl2}
\bibfield{author}{\bibinfo{person}{Seth~D. Baum}, \bibinfo{person}{Ben
  Goertzel}, {and} \bibinfo{person}{Ted~G. Goertzel}.}
  \bibinfo{year}{2011}\natexlab{}.
\newblock \showarticletitle{{How long until human-level AI? Results from an
  expert assessment}}.
\newblock \bibinfo{journal}{\emph{Technological Forecasting \& Social Change}}
  \bibinfo{volume}{78} (\bibinfo{year}{2011}), \bibinfo{pages}{185--195}.
\newblock
Issue 1.


\bibitem[\protect\citeauthoryear{{Berger, Guy}}{{Berger, Guy}}{2018}]%
        {LinkedInEmergingJobs}
\bibfield{author}{\bibinfo{person}{{Berger, Guy}}.}
  \bibinfo{year}{2018}\natexlab{}.
\newblock \bibinfo{title}{{LinkedIn 2018 Emerging Jobs Report}}.
\newblock
\newblock
\urldef\tempurl%
\url{https://economicgraph.linkedin.com/en-us/research/linkedin-2018-emerging-jobs-report}
\showURL{%
\tempurl}
\newblock
\shownote{Accessed January 17, 2021.}


\bibitem[\protect\citeauthoryear{Bonnell, Borda, Daniilidis, Emanuel, Gadsden,
  Gans, Goldberg, Horowitz, Kahn, Kumar, Moore, Moore, Parmacek, Roth, Ruhl,
  Schnall, and Shi}{Bonnell et~al\mbox{.}}{2020}]%
        {PennNSCAI}
\bibfield{author}{\bibinfo{person}{Dawn~A. Bonnell},
  \bibinfo{person}{Michael~J. Borda}, \bibinfo{person}{Kostas Daniilidis},
  \bibinfo{person}{Ezekiel~J. Emanuel}, \bibinfo{person}{Amy~E. Gadsden},
  \bibinfo{person}{John Gans}, \bibinfo{person}{Lee~R. Goldberg},
  \bibinfo{person}{Michael~C. Horowitz}, \bibinfo{person}{Lauren Kahn},
  \bibinfo{person}{Vijay Kumar}, \bibinfo{person}{Jason~H. Moore},
  \bibinfo{person}{Scott Moore}, \bibinfo{person}{Michael~S. Parmacek},
  \bibinfo{person}{Dan Roth}, \bibinfo{person}{Christian Ruhl},
  \bibinfo{person}{Mitchell~D. Schnall}, {and} \bibinfo{person}{Jianbo Shi}.}
  \bibinfo{year}{2020}\natexlab{}.
\newblock \bibinfo{booktitle}{\emph{{University of Pennsylvania Input to
  National Security Commission on Artificial Intelligence}}}.
\newblock \bibinfo{type}{White Paper}. \bibinfo{institution}{Perry World
  House}.
\newblock
\urldef\tempurl%
\url{https://drive.google.com/file/d/1L587pZsBvxssQZljjOxp9cdBvXCvel8N/view}
\showURL{%
\tempurl}


\bibitem[\protect\citeauthoryear{Bosetti, Cattaneo, and Verdolini}{Bosetti
  et~al\mbox{.}}{2015}]%
        {BosettiCattaneo}
\bibfield{author}{\bibinfo{person}{Valentina Bosetti},
  \bibinfo{person}{Cristina Cattaneo}, {and} \bibinfo{person}{Elena
  Verdolini}.} \bibinfo{year}{2015}\natexlab{}.
\newblock \showarticletitle{{Migration of skilled workers and innovation: A
  European perspective}}.
\newblock \bibinfo{journal}{\emph{Journal of International Economics}}
  \bibinfo{volume}{96} (\bibinfo{year}{2015}), \bibinfo{pages}{311--322}.
\newblock
Issue 2.
\urldef\tempurl%
\url{https://EconPapers.repec.org/RePEc:eee:inecon:v:96:y:2015:i:2:p:311-322}
\showURL{%
\tempurl}


\bibitem[\protect\citeauthoryear{Brown, Earle, Kim, and Lee}{Brown
  et~al\mbox{.}}{2013}]%
        {BrownEarleKim}
\bibfield{author}{\bibinfo{person}{J.~David Brown}, \bibinfo{person}{John~S.
  Earle}, \bibinfo{person}{Mee~Jung Kim}, {and} \bibinfo{person}{Kyung~Min
  Lee}.} \bibinfo{year}{2013}\natexlab{}.
\newblock \showarticletitle{{Immigrant entrepreneurs and innovation in the US
  high-tech sector}}.
\newblock In \bibinfo{booktitle}{\emph{The Roles of Immigrants and Foreign
  Students in US Science, Innovation, and Entrepreneurship}}.
  \bibinfo{publisher}{University of Chicago Press}, \bibinfo{address}{Chicago}.
\newblock
\showISBNx{978-0-226-69562-4}
\urldef\tempurl%
\url{http://www.nber.org/chapters/c14103}
\showURL{%
\tempurl}


\bibitem[\protect\citeauthoryear{Cameron}{Cameron}{1996}]%
        {Cameron1996}
\bibfield{author}{\bibinfo{person}{G. Cameron}.}
  \bibinfo{year}{1996}\natexlab{}.
\newblock \bibinfo{booktitle}{\emph{{Innovation and Economic Growth}}}.
\newblock \bibinfo{type}{Discussion Paper}. \bibinfo{institution}{Centre for
  Economic Performance, London School of Economics and Political Science}.
\newblock
\urldef\tempurl%
\url{http://eprints.lse.ac.uk/id/eprint/20685}
\showURL{%
\tempurl}


\bibitem[\protect\citeauthoryear{Chuvpilo}{Chuvpilo}{2018}]%
        {Chuvpilo2018}
\bibfield{author}{\bibinfo{person}{Gleb Chuvpilo}.}
  \bibinfo{year}{2018}\natexlab{}.
\newblock \bibinfo{title}{{Who’s Ahead in AI Research? Insights from NIPS,
  Most Prestigious AI Conference}}.
\newblock
\newblock
\urldef\tempurl%
\url{https://chuvpilo.medium.com/whos-ahead-in-ai-research-insights-from-nips-most-prestigious-ai-conference-df2c361236f6}
\showURL{%
\tempurl}
\newblock
\shownote{Accessed January 17, 2021.}


\bibitem[\protect\citeauthoryear{Chuvpilo}{Chuvpilo}{2019}]%
        {Chuvpilo2019}
\bibfield{author}{\bibinfo{person}{Gleb Chuvpilo}.}
  \bibinfo{year}{2019}\natexlab{}.
\newblock \bibinfo{title}{{AI Research Rankings 2019: Insights from NeurIPS and
  ICML, Leading AI Conferences}}.
\newblock
\newblock
\urldef\tempurl%
\url{https://chuvpilo.medium.com/ai-research-rankings-2019-insights-from-neurips-and-icml-leading-ai-conferences-ee6953152c1a}
\showURL{%
\tempurl}
\newblock
\shownote{Accessed January 17, 2021.}


\bibitem[\protect\citeauthoryear{Chuvpilo}{Chuvpilo}{2020}]%
        {2Chuvpilo2020}
\bibfield{author}{\bibinfo{person}{Gleb Chuvpilo}.}
  \bibinfo{year}{2020}\natexlab{}.
\newblock \bibinfo{title}{{AI Research Rankings 2020: Can the United States
  Stay Ahead of China?}}
\newblock
\newblock
\urldef\tempurl%
\url{https://chuvpilo.medium.com/ai-research-rankings-2020-can-the-united-states-stay-ahead-of-china-61cf14b1216}
\showURL{%
\tempurl}
\newblock
\shownote{Accessed January 17, 2021.}


\bibitem[\protect\citeauthoryear{CIFAR}{CIFAR}{2019}]%
        {topAITalentCanada}
\bibfield{author}{\bibinfo{person}{CIFAR}.} \bibinfo{year}{2019}\natexlab{}.
\newblock \bibinfo{title}{{Canada's top international AI talent grows to 80}}.
\newblock
\newblock
\urldef\tempurl%
\url{https://www.newswire.ca/news-releases/canada-s-top-international-ai-talent-grows-to-80-811709991.html}
\showURL{%
\tempurl}
\newblock
\shownote{Accessed January 17, 2021.}


\bibitem[\protect\citeauthoryear{CIFAR}{CIFAR}{2021}]%
        {Pan-CanadianAIStrategy}
\bibfield{author}{\bibinfo{person}{CIFAR}.} \bibinfo{year}{2021}\natexlab{}.
\newblock \bibinfo{title}{{Pan-Canadian AI Strategy}}.
\newblock
\newblock
\urldef\tempurl%
\url{https://cifar.ca/ai/}
\showURL{%
\tempurl}
\newblock
\shownote{Accessed January 17, 2021.}


\bibitem[\protect\citeauthoryear{{Data USA}}{{Data USA}}{2021}]%
        {AIDataVisualization}
\bibfield{author}{\bibinfo{person}{{Data USA}}.}
  \bibinfo{year}{2021}\natexlab{}.
\newblock \bibinfo{title}{{Artificial Intelligence}}.
\newblock
\newblock
\urldef\tempurl%
\url{https://datausa.io/profile/cip/artificial-intelligence#demographics}
\showURL{%
\tempurl}
\newblock
\shownote{Accessed January 17, 2021.}


\bibitem[\protect\citeauthoryear{Dutton, Barron, and Boskovic}{Dutton
  et~al\mbox{.}}{2018}]%
        {CIFAR2018}
\bibfield{author}{\bibinfo{person}{Tim Dutton}, \bibinfo{person}{Brent Barron},
  {and} \bibinfo{person}{Gaga Boskovic}.} \bibinfo{year}{2018}\natexlab{}.
\newblock \bibinfo{booktitle}{\emph{{Building an AI World: Report on National
  and Regional AI Strategies}}}.
\newblock \bibinfo{type}{{T}echnical {R}eport}. \bibinfo{institution}{CIFAR}.
\newblock
\urldef\tempurl%
\url{https://cifar.ca/wp-content/uploads/2020/05/buildinganaiworld_eng.pdf}
\showURL{%
\tempurl}


\bibitem[\protect\citeauthoryear{Etzioni}{Etzioni}{2019}]%
        {Etzioni2019}
\bibfield{author}{\bibinfo{person}{Oren Etzioni}.}
  \bibinfo{year}{2019}\natexlab{}.
\newblock \showarticletitle{{What Trump’s Executive Order on AI Is Missing}}.
\newblock \bibinfo{journal}{\emph{WIRED}} (\bibinfo{year}{2019}).
\newblock
\urldef\tempurl%
\url{https://www.wired.com/story/what-trumps-executive-order-on-ai-is-missing/}
\showURL{%
\tempurl}
\newblock
\shownote{Accessed January 21, 2021.}


\bibitem[\protect\citeauthoryear{{European Political Strategy
  Centre}}{{European Political Strategy Centre}}{2018}]%
        {EuroCommission}
\bibfield{author}{\bibinfo{person}{{European Political Strategy Centre}}.}
  \bibinfo{year}{2018}\natexlab{}.
\newblock \bibinfo{booktitle}{\emph{{The Age of Artificial Intelligence:
  Towards a European Strategy for Human-Centric Machines}}}.
\newblock \bibinfo{type}{EPSC Strategic Notes} Issue 29.
  \bibinfo{institution}{European Commission}.
\newblock
\urldef\tempurl%
\url{https://ec.europa.eu/jrc/communities/sites/jrccties/files/epsc_strategicnote_ai.pdf}
\showURL{%
\tempurl}


\bibitem[\protect\citeauthoryear{Fagerberg}{Fagerberg}{2006}]%
        {Fagerberg2006}
\bibfield{author}{\bibinfo{person}{Jan Fagerberg}.}
  \bibinfo{year}{2006}\natexlab{}.
\newblock \showarticletitle{Innovation: A guide to the literature}.
\newblock In \bibinfo{booktitle}{\emph{The Oxford Handbook of Innovation}},
  \bibfield{editor}{\bibinfo{person}{Jan Fagerberg} {and}
  \bibinfo{person}{David~C. Mowery}} (Eds.). \bibinfo{publisher}{Oxford
  University Press}, \bibinfo{address}{Oxford}.
\newblock
\urldef\tempurl%
\url{https://doi.org/10.1093/oxfordhb/9780199286805.003.0001}
\showDOI{\tempurl}


\bibitem[\protect\citeauthoryear{Fairlie, Morelix, and Tareque}{Fairlie
  et~al\mbox{.}}{2017}]%
        {Kauffman2017}
\bibfield{author}{\bibinfo{person}{Robert Fairlie}, \bibinfo{person}{Arnobio
  Morelix}, {and} \bibinfo{person}{Inara Tareque}.}
  \bibinfo{year}{2017}\natexlab{}.
\newblock \bibinfo{booktitle}{\emph{{Startup Activity: National Trends}}}.
\newblock \bibinfo{type}{The Kauffman Index}. \bibinfo{institution}{Ewing
  Marion Kauffman Foundation}.
\newblock
\urldef\tempurl%
\url{https://www.kauffman.org/wp-content/uploads/2019/09/2017_Kauffman_Index_Startup_Activity_National_Report_Final.pdf}
\showURL{%
\tempurl}


\bibitem[\protect\citeauthoryear{Gehlhaus and Mutis}{Gehlhaus and
  Mutis}{2021}]%
        {GehlhausMutis2021}
\bibfield{author}{\bibinfo{person}{Diana Gehlhaus} {and}
  \bibinfo{person}{Santiago Mutis}.} \bibinfo{year}{2021}\natexlab{}.
\newblock \bibinfo{booktitle}{\emph{{The U.S. AI Workforce: Understanding the
  Supply of AI Talent}}}.
\newblock \bibinfo{type}{{T}echnical {R}eport}. \bibinfo{institution}{Center
  for Security and Emerging Technology}.
\newblock
\urldef\tempurl%
\url{https://cset.georgetown.edu/research/the-u-s-ai-workforce/}
\showURL{%
\tempurl}


\bibitem[\protect\citeauthoryear{GOV.UK}{GOV.UK}{2020}]%
        {Gov.uk}
\bibfield{author}{\bibinfo{person}{GOV.UK}.} \bibinfo{year}{2020}\natexlab{}.
\newblock \bibinfo{title}{{Home Secretary announces new UK points-based
  immigration system}}.
\newblock
\newblock
\urldef\tempurl%
\url{https://www.gov.uk/government/news/home-secretary-announces-new-uk-points-based-immigration-system}
\showURL{%
\tempurl}
\newblock
\shownote{Accessed January 17, 2021.}


\bibitem[\protect\citeauthoryear{Grace, Salvatier, Dafoe, Zhang, and
  Evans}{Grace et~al\mbox{.}}{2018}]%
        {GraceEtAl}
\bibfield{author}{\bibinfo{person}{Katja Grace}, \bibinfo{person}{John
  Salvatier}, \bibinfo{person}{Allan Dafoe}, \bibinfo{person}{Baobao Zhang},
  {and} \bibinfo{person}{Owain Evans}.} \bibinfo{year}{2018}\natexlab{}.
\newblock \showarticletitle{{When will AI exceed human performance? Evidence
  from AI experts}}.
\newblock \bibinfo{journal}{\emph{Journal of Artificial Intelligence Research}}
   \bibinfo{volume}{62} (\bibinfo{year}{2018}), \bibinfo{pages}{729--754}.
\newblock
\urldef\tempurl%
\url{https://doi.org/10.1613/jair.1.11222}
\showURL{%
\tempurl}


\bibitem[\protect\citeauthoryear{Gruetzemacher, Paradice, and
  Lee}{Gruetzemacher et~al\mbox{.}}{2020}]%
        {gruetzemacher2019forecasting}
\bibfield{author}{\bibinfo{person}{Ross Gruetzemacher}, \bibinfo{person}{David
  Paradice}, {and} \bibinfo{person}{Kang~Bok Lee}.}
  \bibinfo{year}{2020}\natexlab{}.
\newblock \showarticletitle{{Forecasting extreme labor displacement: A survey
  of AI practitioners}}.
\newblock \bibinfo{journal}{\emph{Technological Forecasting and Social Change}}
   \bibinfo{volume}{161} (\bibinfo{year}{2020}).
\newblock
\newblock
\shownote{120323.}


\bibitem[\protect\citeauthoryear{Han and Appelbaum}{Han and Appelbaum}{2016}]%
        {HanAppelbaum}
\bibfield{author}{\bibinfo{person}{Xueying Han} {and}
  \bibinfo{person}{Richard~P. Appelbaum}.} \bibinfo{year}{2016}\natexlab{}.
\newblock \bibinfo{booktitle}{\emph{{Will They Stay or Will They Go?:
  International STEM Students Are up for Grabs}}}.
\newblock \bibinfo{type}{{T}echnical {R}eport}. \bibinfo{institution}{Ewing
  Marion Kauffman Foundation}.
\newblock
\urldef\tempurl%
\url{https://eric.ed.gov/?id=ED570660}
\showURL{%
\tempurl}


\bibitem[\protect\citeauthoryear{Hao}{Hao}{2019}]%
        {Hao2019}
\bibfield{author}{\bibinfo{person}{Karen Hao}.}
  \bibinfo{year}{2019}\natexlab{}.
\newblock \showarticletitle{{Yes, China is probably outspending the US in
  AI—but not on defense}}.
\newblock \bibinfo{journal}{\emph{MIT Technology Review}}
  (\bibinfo{year}{2019}).
\newblock
\urldef\tempurl%
\url{https://www.technologyreview.com/2019/12/05/65019/china-us-ai-military-spending/}
\showURL{%
\tempurl}
\newblock
\shownote{Accessed January 21, 2021.}


\bibitem[\protect\citeauthoryear{Hao}{Hao}{2020}]%
        {Hao2020}
\bibfield{author}{\bibinfo{person}{Karen Hao}.}
  \bibinfo{year}{2020}\natexlab{}.
\newblock \showarticletitle{{Trump’s freeze on new visas could threaten US
  dominance in AI}}.
\newblock \bibinfo{journal}{\emph{MIT Technology Review}}
  (\bibinfo{year}{2020}).
\newblock
\urldef\tempurl%
\url{https://www.technologyreview.com/2020/06/26/1004520/trump-executive-order-h1b-visa-threatens-us-ai/}
\showURL{%
\tempurl}
\newblock
\shownote{Accessed January 27, 2021.}


\bibitem[\protect\citeauthoryear{Huang and Arnold}{Huang and Arnold}{2020}]%
        {HuangArnold2020}
\bibfield{author}{\bibinfo{person}{Tina Huang} {and} \bibinfo{person}{Zachary
  Arnold}.} \bibinfo{year}{2020}\natexlab{}.
\newblock \bibinfo{booktitle}{\emph{{Immigration Policy and the Global
  Competition for AI Talent}}}.
\newblock \bibinfo{type}{{T}echnical {R}eport}. \bibinfo{institution}{Center
  for Security and Emerging Technology}.
\newblock
\urldef\tempurl%
\url{https://cset.georgetown.edu/research/immigration-policy-and-the-global-competition-for-ai-talent/}
\showURL{%
\tempurl}


\bibitem[\protect\citeauthoryear{Huang, Arnold, and Zwetsloot}{Huang
  et~al\mbox{.}}{2020}]%
        {HuangArnoldZwetsloot2020}
\bibfield{author}{\bibinfo{person}{Tina Huang}, \bibinfo{person}{Zachary
  Arnold}, {and} \bibinfo{person}{Remco Zwetsloot}.}
  \bibinfo{year}{2020}\natexlab{}.
\newblock \bibinfo{booktitle}{\emph{{Most of America’s ``Most Promising'' AI
  Startups Have Immigrant Founders}}}.
\newblock \bibinfo{type}{Data Brief}. \bibinfo{institution}{Center for Security
  and Emerging Technology}.
\newblock
\urldef\tempurl%
\url{https://cset.georgetown.edu/research/most-of-americas-most-promising-ai-startups-have-immigrant-founders/}
\showURL{%
\tempurl}


\bibitem[\protect\citeauthoryear{Hunt and Gauthier-Loiselle}{Hunt and
  Gauthier-Loiselle}{2010}]%
        {HuntGauthier-Loiselle}
\bibfield{author}{\bibinfo{person}{Jennifer Hunt} {and}
  \bibinfo{person}{Marjolaine Gauthier-Loiselle}.}
  \bibinfo{year}{2010}\natexlab{}.
\newblock \showarticletitle{{How much does immigration boost innovation?}}
\newblock \bibinfo{journal}{\emph{American Economic Journal: Macroeconomics}}
  \bibinfo{volume}{2} (\bibinfo{year}{2010}), \bibinfo{pages}{31--56}.
\newblock
Issue 2.
\urldef\tempurl%
\url{www.jstor.org/stable/25760296}
\showURL{%
\tempurl}


\bibitem[\protect\citeauthoryear{{JF Gagne}}{{JF Gagne}}{2019}]%
        {Gagne2019}
\bibfield{author}{\bibinfo{person}{{JF Gagne}}.}
  \bibinfo{year}{2019}\natexlab{}.
\newblock \bibinfo{title}{{Global AI Talent Report 2019}}.
\newblock
\newblock
\urldef\tempurl%
\url{https://jfgagne.ai/talent-2019/}
\showURL{%
\tempurl}
\newblock
\shownote{Accessed January 17, 2021.}


\bibitem[\protect\citeauthoryear{Jobin, Ienca, and Vayena}{Jobin
  et~al\mbox{.}}{2019}]%
        {Jobinetal2019}
\bibfield{author}{\bibinfo{person}{Anna Jobin}, \bibinfo{person}{Marcello
  Ienca}, {and} \bibinfo{person}{Effy Vayena}.}
  \bibinfo{year}{2019}\natexlab{}.
\newblock \showarticletitle{{The global landscape of AI ethics guidelines}}.
\newblock \bibinfo{journal}{\emph{Nature}} (\bibinfo{year}{2019}).
\newblock
\urldef\tempurl%
\url{https://www.nature.com/articles/s42256-019-0088-2}
\showURL{%
\tempurl}
\newblock
\shownote{Accessed January 27, 2021.}


\bibitem[\protect\citeauthoryear{Kerr, Kerr, \"{O}zden, and Parsons}{Kerr
  et~al\mbox{.}}{2017}]%
        {KerrOzden2017}
\bibfield{author}{\bibinfo{person}{Sari~Pekkala Kerr}, \bibinfo{person}{William
  Kerr}, \bibinfo{person}{\c{C}a\u{g}lar \"{O}zden}, {and}
  \bibinfo{person}{Christopher Parsons}.} \bibinfo{year}{2017}\natexlab{}.
\newblock \showarticletitle{{High-skilled migration and agglomeration}}.
\newblock \bibinfo{journal}{\emph{Annual Review of Economics}}
  \bibinfo{volume}{9} (\bibinfo{year}{2017}), \bibinfo{pages}{201--234}.
\newblock
\urldef\tempurl%
\url{https://doi.org/10.1146/annurev-economics-063016-103705}
\showURL{%
\tempurl}


\bibitem[\protect\citeauthoryear{Kerr}{Kerr}{2019}]%
        {Kerr2019}
\bibfield{author}{\bibinfo{person}{William~R. Kerr}.}
  \bibinfo{year}{2019}\natexlab{}.
\newblock \bibinfo{booktitle}{\emph{The Gift of Global Talent: How Migration
  Shapes Business, Economy \& Society}}.
\newblock \bibinfo{publisher}{Stanford University Press},
  \bibinfo{address}{Palo Alto, CA}.
\newblock


\bibitem[\protect\citeauthoryear{Krafft, Young, Katell, Huang, and
  Bugingo}{Krafft et~al\mbox{.}}{2020}]%
        {krafft2020defining}
\bibfield{author}{\bibinfo{person}{P.~M. Krafft}, \bibinfo{person}{Meg Young},
  \bibinfo{person}{Michael Katell}, \bibinfo{person}{Karen Huang}, {and}
  \bibinfo{person}{Ghislain Bugingo}.} \bibinfo{year}{2020}\natexlab{}.
\newblock \showarticletitle{{Defining AI in policy versus practice}}. In
  \bibinfo{booktitle}{\emph{Proceedings of the AAAI/ACM Conference on AI,
  Ethics, and Society}} (New York, NY, USA) \emph{(\bibinfo{series}{AIES
  ’20})}. \bibinfo{publisher}{Association for Computing Machinery},
  \bibinfo{address}{New York, NY, USA}, \bibinfo{pages}{72–78}.
\newblock
\showISBNx{9781450371100}
\urldef\tempurl%
\url{https://doi.org/10.1145/3375627.3375835}
\showDOI{\tempurl}


\bibitem[\protect\citeauthoryear{Leopold, Ratcheva, and Zahidi}{Leopold
  et~al\mbox{.}}{2018}]%
        {WEFJobReport}
\bibfield{author}{\bibinfo{person}{Till~Alexander Leopold},
  \bibinfo{person}{Vesselina Ratcheva}, {and} \bibinfo{person}{Saadia Zahidi}.}
  \bibinfo{year}{2018}\natexlab{}.
\newblock \bibinfo{booktitle}{\emph{{The Future of Jobs Report 2018}}}.
\newblock \bibinfo{type}{Insight Report}. \bibinfo{institution}{World Economic
  Forum}.
\newblock
\urldef\tempurl%
\url{http://www3.weforum.org/docs/WEF_Future_of_Jobs_2018.pdf}
\showURL{%
\tempurl}


\bibitem[\protect\citeauthoryear{Markow, Hughes, and Bundy}{Markow
  et~al\mbox{.}}{2018}]%
        {Markow2018}
\bibfield{author}{\bibinfo{person}{Will Markow}, \bibinfo{person}{Debbie
  Hughes}, {and} \bibinfo{person}{Andrew Bundy}.}
  \bibinfo{year}{2018}\natexlab{}.
\newblock \bibinfo{booktitle}{\emph{{The New Foundational Skills of the Digital
  Economy: Developing the Professionals of the Future}}}.
\newblock \bibinfo{type}{{T}echnical {R}eport}. \bibinfo{institution}{Burning
  Glass / BHEF}.
\newblock
\urldef\tempurl%
\url{https://www.bhef.com/sites/default/files/BHEF_2018_New_Foundational_Skills.pdf}
\showURL{%
\tempurl}


\bibitem[\protect\citeauthoryear{M\"{u}ller and Bostrom}{M\"{u}ller and
  Bostrom}{2016}]%
        {MullerBostrom}
\bibfield{author}{\bibinfo{person}{Vincent~C. M\"{u}ller} {and}
  \bibinfo{person}{Nick Bostrom}.} \bibinfo{year}{2016}\natexlab{}.
\newblock \showarticletitle{Future progress in artificial intelligence: A
  survey of expert opinion}.
\newblock In \bibinfo{booktitle}{\emph{Fundamental Issues of Artificial
  Intelligence}}, \bibfield{editor}{\bibinfo{person}{Vincent M\"{u}ller}}
  (Ed.). \bibinfo{publisher}{Springer International Publishing},
  \bibinfo{address}{Cham}, \bibinfo{pages}{553--571}.
\newblock
\showISBNx{978-3-319-26483-7}
\urldef\tempurl%
\url{https://doi.org/10.1007/978-3-319-26485-1_33}
\showDOI{\tempurl}


\bibitem[\protect\citeauthoryear{Musumba, Jin, and Mjelde}{Musumba
  et~al\mbox{.}}{2011}]%
        {MusumbaJinMjelde}
\bibfield{author}{\bibinfo{person}{Mark Musumba}, \bibinfo{person}{Yanhong~H.
  Jin}, {and} \bibinfo{person}{James~W. Mjelde}.}
  \bibinfo{year}{2011}\natexlab{}.
\newblock \showarticletitle{Factors influencing career location preferences of
  international graduate students in the United States}.
\newblock \bibinfo{journal}{\emph{Education Economics}} \bibinfo{volume}{19},
  \bibinfo{number}{5} (\bibinfo{year}{2011}), \bibinfo{pages}{501--517}.
\newblock
\urldef\tempurl%
\url{https://doi.org/10.1080/09645290903102902}
\showDOI{\tempurl}


\bibitem[\protect\citeauthoryear{Ozgen, Nijkamp, and Poot}{Ozgen
  et~al\mbox{.}}{2011}]%
        {OzgenNijkamp}
\bibfield{author}{\bibinfo{person}{Ceren Ozgen}, \bibinfo{person}{Peter
  Nijkamp}, {and} \bibinfo{person}{Jacques Poot}.}
  \bibinfo{year}{2011}\natexlab{}.
\newblock \bibinfo{booktitle}{\emph{{Immigration and Innovation in European
  Regions}}}.
\newblock \bibinfo{type}{TI Discussion Papers} 11-112/3.
  \bibinfo{institution}{Tinbergen Institute}.
\newblock
\urldef\tempurl%
\url{http://www.tinbergen.nl/ti-publications/discussion-papers.php?paper=1803}
\showURL{%
\tempurl}


\bibitem[\protect\citeauthoryear{Perrault, Shoham, Brynjolfsson, Clark,
  Etchemendy, Grosz, Lyons, Manyika, Mishra, and Niebles}{Perrault
  et~al\mbox{.}}{2019}]%
        {PerraultEtAl2019}
\bibfield{author}{\bibinfo{person}{Raymond Perrault}, \bibinfo{person}{Yoav
  Shoham}, \bibinfo{person}{Erik Brynjolfsson}, \bibinfo{person}{Jack Clark},
  \bibinfo{person}{John Etchemendy}, \bibinfo{person}{Barbara Grosz},
  \bibinfo{person}{Terah Lyons}, \bibinfo{person}{James Manyika},
  \bibinfo{person}{Saurabh Mishra}, {and} \bibinfo{person}{Juan~Carlos
  Niebles}.} \bibinfo{year}{2019}\natexlab{}.
\newblock \bibinfo{booktitle}{\emph{{The AI Index 2019 Annual Report}}}.
\newblock \bibinfo{type}{{T}echnical {R}eport}. \bibinfo{institution}{AI Index
  Steering Committee, Human-Centered AI Institute, Stanford University}.
\newblock
\urldef\tempurl%
\url{https://hai.stanford.edu/research/ai-index-2019}
\showURL{%
\tempurl}


\bibitem[\protect\citeauthoryear{Roberts, Cowls, Morley, Taddeo, Wang, and
  Floridi}{Roberts et~al\mbox{.}}{2020}]%
        {RobertsCowlsMorley}
\bibfield{author}{\bibinfo{person}{Huw Roberts}, \bibinfo{person}{Josh Cowls},
  \bibinfo{person}{Jessica Morley}, \bibinfo{person}{Mariarosaria Taddeo},
  \bibinfo{person}{Vincent Wang}, {and} \bibinfo{person}{Luciano Floridi}.}
  \bibinfo{year}{2020}\natexlab{}.
\newblock \showarticletitle{{The Chinese approach to artificial intelligence:
  an analysis of policy, ethics, and regulation}}.
\newblock \bibinfo{journal}{\emph{AI \& Society}} (\bibinfo{year}{2020}).
\newblock
\urldef\tempurl%
\url{https://doi.org/10.1007/s00146-020-00992-2}
\showURL{%
\tempurl}


\bibitem[\protect\citeauthoryear{Roca}{Roca}{2019}]%
        {Roca2019}
\bibfield{author}{\bibinfo{person}{Thomas Roca}.}
  \bibinfo{year}{2019}\natexlab{}.
\newblock \bibinfo{booktitle}{\emph{{Identifying AI talents among LinkedIn
  members: A machine learning approach}}}.
\newblock \bibinfo{type}{LinkedIn Economic Graph}.
  \bibinfo{institution}{LinkedIn}.
\newblock
\urldef\tempurl%
\url{https://aiforall.azurewebsites.net/pdf/AI%20in%20the%20Labour%20Force%20to%20share.pdf}
\showURL{%
\tempurl}


\bibitem[\protect\citeauthoryear{Sandberg and Bostrom}{Sandberg and
  Bostrom}{2011}]%
        {SandbergBostrom2011}
\bibfield{author}{\bibinfo{person}{Anders Sandberg} {and} \bibinfo{person}{Nick
  Bostrom}.} \bibinfo{year}{2011}\natexlab{}.
\newblock \bibinfo{booktitle}{\emph{Machine Intelligence Survey}}.
\newblock \bibinfo{type}{{T}echnical {R}eport}. \bibinfo{institution}{Future of
  Humanity Institute, Oxford University}. \bibinfo{pages}{1--12} pages.
\newblock
\urldef\tempurl%
\url{https://www.fhi.ox.ac.uk/reports/2011-1.pdf}
\showURL{%
\tempurl}
\newblock
\shownote{Technical Report \#2011-1.}


\bibitem[\protect\citeauthoryear{Savage}{Savage}{2020}]%
        {Savage2020}
\bibfield{author}{\bibinfo{person}{Neil Savage}.}
  \bibinfo{year}{2020}\natexlab{}.
\newblock \showarticletitle{{Learning the algorithms of power}}.
\newblock \bibinfo{journal}{\emph{Nature}}  \bibinfo{volume}{588}
  (\bibinfo{year}{2020}), \bibinfo{pages}{s102--s104}.
\newblock
\urldef\tempurl%
\url{https://www.nature.com/articles/d41586-020-03409-8}
\showURL{%
\tempurl}


\bibitem[\protect\citeauthoryear{Saxenian}{Saxenian}{1999}]%
        {Saxenian1999}
\bibfield{author}{\bibinfo{person}{Annalee Saxenian}.}
  \bibinfo{year}{1999}\natexlab{}.
\newblock \bibinfo{booktitle}{\emph{{Silicon Valley’s New Immigrant
  Entrepreneurs}}}.
\newblock \bibinfo{type}{{T}echnical {R}eport}. \bibinfo{institution}{Public
  Policy Institute of California}.
\newblock
\urldef\tempurl%
\url{https://www.ppic.org/publication/silicon-valleys-new-immigrant-entrepreneurs/}
\showURL{%
\tempurl}


\bibitem[\protect\citeauthoryear{Schmidt, Work, Catz, Chien, Clyburn, Darby,
  Ford, Griffiths, Horvitz, Jassy, Louie, Mark, Matheny, McFarland, and
  Moore}{Schmidt et~al\mbox{.}}{2020}]%
        {NSCAI}
\bibfield{author}{\bibinfo{person}{Eric Schmidt}, \bibinfo{person}{Robert~O.
  Work}, \bibinfo{person}{Safra Catz}, \bibinfo{person}{Steve Chien},
  \bibinfo{person}{Mignon Clyburn}, \bibinfo{person}{Christopher Darby},
  \bibinfo{person}{Kenneth Ford}, \bibinfo{person}{Jos\'{e}-Marie Griffiths},
  \bibinfo{person}{Eric Horvitz}, \bibinfo{person}{Andrew Jassy},
  \bibinfo{person}{Gilman Louie}, \bibinfo{person}{William Mark},
  \bibinfo{person}{Jason Matheny}, \bibinfo{person}{Katharina McFarland}, {and}
  \bibinfo{person}{Andrew Moore}.} \bibinfo{year}{2020}\natexlab{}.
\newblock \bibinfo{booktitle}{\emph{{NSCAI Interim Report and Third Quarter
  Recommendations Memo}}}.
\newblock \bibinfo{type}{Report}. \bibinfo{institution}{National Security
  Commission on Artificial Intelligence}.
\newblock
\urldef\tempurl%
\url{https://www.nscai.gov/reports}
\showURL{%
\tempurl}


\bibitem[\protect\citeauthoryear{Semuels}{Semuels}{2019}]%
        {Semuels2019}
\bibfield{author}{\bibinfo{person}{Alana Semuels}.}
  \bibinfo{year}{2019}\natexlab{}.
\newblock \showarticletitle{{Tech Companies Say it's Too Hard to Hire
  High-Skilled Immigrants in the U.S. -- So They're Growing in Canada
  Instead}}.
\newblock \bibinfo{journal}{\emph{TIME}} (\bibinfo{year}{2019}).
\newblock
\urldef\tempurl%
\url{https://time.com/5634351/canada-high-skilled-labor-immigrants/}
\showURL{%
\tempurl}
\newblock
\shownote{Accessed January 21, 2021.}


\bibitem[\protect\citeauthoryear{Sheehan, Banerjee, Cantara, Kim, and
  Roche}{Sheehan et~al\mbox{.}}{2019}]%
        {GlobalTalentTracker}
\bibfield{author}{\bibinfo{person}{Matt Sheehan}, \bibinfo{person}{Ishan
  Banerjee}, \bibinfo{person}{Annie Cantara}, \bibinfo{person}{Young Kim},
  {and} \bibinfo{person}{Chris Roche}.} \bibinfo{year}{2019}\natexlab{}.
\newblock \bibinfo{title}{{The Global AI Talent Tracker}}.
\newblock
\newblock
\urldef\tempurl%
\url{https://macropolo.org/digital-projects/the-global-ai-talent-tracker/}
\showURL{%
\tempurl}
\newblock
\shownote{Accessed January 17, 2021.}


\bibitem[\protect\citeauthoryear{{Social Sciences and Humanities Research
  Council of Canada}}{{Social Sciences and Humanities Research Council of
  Canada}}{2020}]%
        {CanadaRecruitment2020}
\bibfield{author}{\bibinfo{person}{{Social Sciences and Humanities Research
  Council of Canada}}.} \bibinfo{year}{2020}\natexlab{}.
\newblock \bibinfo{title}{{Government of Canada recruits world-renowned
  researcher to bridge the gap between AI and Neuroscience}}.
\newblock
\newblock
\urldef\tempurl%
\url{https://www.newswire.ca/news-releases/government-of-canada-recruits-world-renowned-researcher-to-bridge-the-gap-between-ai-and-neuroscience-870757268.html}
\showURL{%
\tempurl}
\newblock
\shownote{Accessed January 17, 2021.}


\bibitem[\protect\citeauthoryear{Stephan}{Stephan}{2010}]%
        {Stephan2010}
\bibfield{author}{\bibinfo{person}{Paula~E. Stephan}.}
  \bibinfo{year}{2010}\natexlab{}.
\newblock \showarticletitle{The ``I''s have it: Immigration and innovation, the
  perspective from academe}.
\newblock In \bibinfo{booktitle}{\emph{Innovation Policy and the Economy}},
  \bibfield{editor}{\bibinfo{person}{Josh Lerner} {and} \bibinfo{person}{Scott
  Stern}} (Eds.). Vol.~\bibinfo{volume}{10}. \bibinfo{publisher}{University of
  Chicago Press}, \bibinfo{address}{Chicago}, \bibinfo{pages}{83--127}.
\newblock
\showISBNx{0-226-47334-1}
\urldef\tempurl%
\url{http://www.nber.org/chapters/c11766}
\showURL{%
\tempurl}


\bibitem[\protect\citeauthoryear{{Team YS}}{{Team YS}}{2017}]%
        {TeamYS}
\bibfield{author}{\bibinfo{person}{{Team YS}}.}
  \bibinfo{year}{2017}\natexlab{}.
\newblock \showarticletitle{{Now, France has a visa for startup entrepreneurs,
  techies, and VCs}}.
\newblock \bibinfo{journal}{\emph{Your Story}} (\bibinfo{year}{2017}).
\newblock
\urldef\tempurl%
\url{https://yourstory.com/2017/01/french-tech-visa?utm_pageloadtype=scroll}
\showURL{%
\tempurl}
\newblock
\shownote{Accessed January 21, 2021.}


\bibitem[\protect\citeauthoryear{Toney and Flagg}{Toney and Flagg}{2020}]%
        {ToneyFlagg2020}
\bibfield{author}{\bibinfo{person}{Autumn Toney} {and} \bibinfo{person}{Melissa
  Flagg}.} \bibinfo{year}{2020}\natexlab{}.
\newblock \bibinfo{booktitle}{\emph{{U.S. Demand for AI-Related Talent}}}.
\newblock \bibinfo{type}{Data Brief}. \bibinfo{institution}{Center for Security
  and Emerging Technology}.
\newblock
\urldef\tempurl%
\url{https://cset.georgetown.edu/research/u-s-demand-for-ai-related-talent/}
\showURL{%
\tempurl}


\bibitem[\protect\citeauthoryear{Van~Noorden}{Van~Noorden}{2012}]%
        {VanNoorden2012}
\bibfield{author}{\bibinfo{person}{Richard Van~Noorden}.}
  \bibinfo{year}{2012}\natexlab{}.
\newblock \showarticletitle{{Global mobility: Science on the move}}.
\newblock \bibinfo{journal}{\emph{Nature News}}  \bibinfo{volume}{490}
  (\bibinfo{year}{2012}), \bibinfo{pages}{326–329}.
\newblock
\urldef\tempurl%
\url{https://www.nature.com/news/global-mobility-science-on-the-move-1.11602}
\showURL{%
\tempurl}
\newblock
\shownote{Accessed January 17, 2021.}


\bibitem[\protect\citeauthoryear{{Venture Scanner}}{{Venture Scanner}}{2017}]%
        {VentureScanner}
\bibfield{author}{\bibinfo{person}{{Venture Scanner}}.}
  \bibinfo{year}{2017}\natexlab{}.
\newblock \bibinfo{title}{{Artificial Intelligence Startup Highlights – Q4
  2017}}.
\newblock
\newblock
\urldef\tempurl%
\url{https://www.venturescanner.com/2018/01/11/artificial-intelligence-startup-highlights-q4-2017/}
\showURL{%
\tempurl}
\newblock
\shownote{Accessed January 17, 2021.}


\bibitem[\protect\citeauthoryear{Verspagen}{Verspagen}{2006}]%
        {Verspagen2006}
\bibfield{author}{\bibinfo{person}{Bart Verspagen}.}
  \bibinfo{year}{2006}\natexlab{}.
\newblock \showarticletitle{Innovation and economic growth}.
\newblock In \bibinfo{booktitle}{\emph{The Oxford Handbook of Innovation}},
  \bibfield{editor}{\bibinfo{person}{Jan Fagerberg} {and}
  \bibinfo{person}{David~C. Mowery}} (Eds.). \bibinfo{publisher}{Oxford
  University Press}, \bibinfo{address}{Oxford}.
\newblock
\urldef\tempurl%
\url{https://doi.org/10.1093/oxfordhb/9780199286805.003.0018}
\showDOI{\tempurl}


\bibitem[\protect\citeauthoryear{Wadhwa, Saxenian, Rissing, and Gereffi}{Wadhwa
  et~al\mbox{.}}{2007}]%
        {WadhwaSaxenian}
\bibfield{author}{\bibinfo{person}{Vivek Wadhwa}, \bibinfo{person}{AnnaLee
  Saxenian}, \bibinfo{person}{Ben~A. Rissing}, {and} \bibinfo{person}{Gary
  Gereffi}.} \bibinfo{year}{2007}\natexlab{}.
\newblock \bibinfo{booktitle}{\emph{{America's New Immigrant Entrepreneurs:
  Part I}}}.
\newblock \bibinfo{type}{Duke Science, Technology \& Innovation Paper}~23.
  \bibinfo{institution}{Duke University}.
\newblock
\urldef\tempurl%
\url{http://dx.doi.org/10.2139/ssrn.990152}
\showURL{%
\tempurl}


\bibitem[\protect\citeauthoryear{Walsh}{Walsh}{2018}]%
        {walsh2018expert}
\bibfield{author}{\bibinfo{person}{Toby Walsh}.}
  \bibinfo{year}{2018}\natexlab{}.
\newblock \showarticletitle{Expert and non-expert opinion about technological
  unemployment}.
\newblock \bibinfo{journal}{\emph{International Journal of Automation and
  Computing}} \bibinfo{volume}{15}, \bibinfo{number}{5} (\bibinfo{year}{2018}),
  \bibinfo{pages}{637--642}.
\newblock


\bibitem[\protect\citeauthoryear{Weinstein}{Weinstein}{2020}]%
        {ChinaTalentTracker}
\bibfield{author}{\bibinfo{person}{Emily Weinstein}.}
  \bibinfo{year}{2020}\natexlab{}.
\newblock \bibinfo{booktitle}{\emph{{Chinese Talent Program Tracker}}}.
\newblock \bibinfo{type}{Data Visualization}. \bibinfo{institution}{Center for
  Security and Emerging Technology}.
\newblock
\urldef\tempurl%
\url{https://cset.georgetown.edu/research/chinese-talent-program-tracker/}
\showURL{%
\tempurl}
\newblock
\shownote{Accessed January 17, 2021.}


\bibitem[\protect\citeauthoryear{Zhang, Anderljung, Kahn, Dreksler, Horowitz,
  and Dafoe}{Zhang et~al\mbox{.}}{2020}]%
        {ZhangAnderljungetal}
\bibfield{author}{\bibinfo{person}{Baobao Zhang}, \bibinfo{person}{Markus
  Anderljung}, \bibinfo{person}{Lauren Kahn}, \bibinfo{person}{Noemi Dreksler},
  \bibinfo{person}{Michael~C. Horowitz}, {and} \bibinfo{person}{Allan Dafoe}.}
  \bibinfo{year}{2020}\natexlab{}.
\newblock \bibinfo{title}{Ethics and governance of artificial intelligence:
  Evidence from a survey of machine learning researchers}.
  (\bibinfo{year}{2020}).
\newblock
\urldef\tempurl%
\url{https://osf.io/pnvfd/}
\showURL{%
\tempurl}
\newblock
\shownote{Preprint.}


\bibitem[\protect\citeauthoryear{Zhang and Dafoe}{Zhang and Dafoe}{2020}]%
        {zhang2020us}
\bibfield{author}{\bibinfo{person}{Baobao Zhang} {and} \bibinfo{person}{Allan
  Dafoe}.} \bibinfo{year}{2020}\natexlab{}.
\newblock \showarticletitle{{US public opinion on the governance of artificial
  intelligence}}. In \bibinfo{booktitle}{\emph{Proceedings of the AAAI/ACM
  Conference on AI, Ethics, and Society}} \emph{(\bibinfo{series}{AIES
  ’20})}. \bibinfo{publisher}{Association for Computing Machinery},
  \bibinfo{address}{New York, NY, USA}, \bibinfo{pages}{187–193}.
\newblock
\showISBNx{9781450371100}
\urldef\tempurl%
\url{https://doi.org/10.1145/3375627.3375827}
\showDOI{\tempurl}


\bibitem[\protect\citeauthoryear{Zwetsloot, Dunham, Arnold, and
  Huang}{Zwetsloot et~al\mbox{.}}{2019a}]%
        {2Zwetsloot2019}
\bibfield{author}{\bibinfo{person}{Remco Zwetsloot}, \bibinfo{person}{James
  Dunham}, \bibinfo{person}{Zachary Arnold}, {and} \bibinfo{person}{Tina
  Huang}.} \bibinfo{year}{2019}\natexlab{a}.
\newblock \bibinfo{booktitle}{\emph{{Keeping Top AI Talent in the United
  States}}}.
\newblock \bibinfo{type}{{T}echnical {R}eport}. \bibinfo{institution}{Center
  for Security and Emerging Technology}.
\newblock
\urldef\tempurl%
\url{https://cset.georgetown.edu/research/keeping-top-ai-talent-in-the-united-states/}
\showURL{%
\tempurl}


\bibitem[\protect\citeauthoryear{Zwetsloot, Heston, and Arnold}{Zwetsloot
  et~al\mbox{.}}{2019b}]%
        {Zwetsloot2019}
\bibfield{author}{\bibinfo{person}{Remco Zwetsloot}, \bibinfo{person}{Roxanne
  Heston}, {and} \bibinfo{person}{Zachary Arnold}.}
  \bibinfo{year}{2019}\natexlab{b}.
\newblock \bibinfo{booktitle}{\emph{{Strengthening the U.S. AI Workforce: A
  Policy and Research Agenda}}}.
\newblock \bibinfo{type}{{T}echnical {R}eport}. \bibinfo{institution}{Center
  for Security and Emerging Technology}.
\newblock
\urldef\tempurl%
\url{https://cset.georgetown.edu/research/strengthening-the-u-s-ai-workforce/}
\showURL{%
\tempurl}


\end{thebibliography}


\appendix
\section{Appendix}

\subsection{Survey Text}

``Of the following countries, which would you have a greater than 25\% likelihood of moving to for work or study in the next three years?'' 

\bigskip
\noindent The answer options were presented in a randomized order:
\begin{itemize}
    \item United States
    \item United Kingdom
    \item China
    \item Australia
    \item Singapore
    \item Japan
    \item Canada
    \item South Korea
    \item Israel
    \item France
    \item Switzerland
    \item Other [text box]
    \item None of these
\end{itemize}
\bigskip
``When considering moving to work or study in a country that you don’t currently work or study in full-time, what factors are important in your consideration?''

\bigskip
\noindent The answer options were presented in a randomized order:
\begin{itemize}
    \item Current immigration difficulties in country of residence
    \item Ease of immigration and/or incentives provided to move to destination country
    \item Personal relations in the destination country (e.g., friends and family)
    \item Good professional opportunities and environment (e.g., you are offered a job at an attractive organization)
    \item Lifestyle and culture
    \item Political climate of the destination country
    \item Other: [text box]
    \item None of the above
\end{itemize}

\bigskip
\noindent ``Which of the following, if any, are serious problems for conducting high-quality AI research in [INSERT NAME OF YOUR COUNTRY OF WORK] today?''

\bigskip
\noindent The answer options were presented in a randomized order:

\begin{itemize}
    \item Lack of government funding for AI research
    \item Lack of corporate funding for AI research
    \item Lack of funding for training students
    \item Visa and immigration problems facing foreign researchers or students who want to work or study in [INSERT NAME OF YOUR COUNTRY OF WORK]
    \item Lack of a successful ecosystem for AI startups
    \item Not enough top researchers and labs
    \item The political climate in [INSERT NAME OF YOUR COUNTRY OF WORK] is not conducive to AI research
    \item Other: [text box]
    \item There are no serious problems
\end{itemize}

\subsection{Sample Representativeness}

In order to assess sample representativeness, we collected demographic data for a sample of 446 non-respondents. Overall, there did not appear to be concerning levels of response bias. However, a multiple regression of the association between demographic variables and whether an individual from the overall sample responded to the survey showed that, compared to non-respondents, respondents had lower h-indexes (which measures the productivity and citation impact of a researcher) and were more likely to work in academia (rather than industry or another kind of workplace). This academic skew appears to be a common response bias in surveys of AI talent \cite{AikenDunham2020}. It is also worth keeping in mind that respondents and non-respondents were predominantly male (91\% and 89\%, respectively). This reflects the continued gender imbalance that exists in the field \cite{PerraultEtAl2019,Gagne2019,Roca2019}. Nevertheless, our survey achieved a higher response rate and more global coverage than other survey studies of AI researchers.

\end{document}